\documentclass[a4paper,fleqn,usenatbib,twocolumn]{mnras}
\usepackage{hyperref}
\usepackage[utf8]{inputenc}
\usepackage{newtxtext,newtxmath}
\usepackage[T1]{fontenc}
\usepackage{ae,aecompl}
\usepackage[]{units}
\usepackage{graphicx}
\usepackage{amssymb}
\usepackage{amsmath}
\usepackage{threeparttable}
\usepackage{bm}
\usepackage{color}
\usepackage{hhline}
\usepackage{subfigure}
\usepackage{lineno}
\title[Assemby bias and galaxy-galaxy lensing]{Can Assembly Bias Explain the Lensing Amplitude of the BOSS CMASS Sample in a Planck Cosmology?}
\author[S. Yuan et al.]{
Sihan Yuan$^{1}$\thanks{E-mail: sihan.yuan@cfa.harvard.edu}, Daniel J. Eisenstein$^{1}$, Alexie Leauthaud$^{2}$
\\
$^{1}$Harvard-Smithsonian Center for Astrophysics, 60 Garden St., Cambridge, MA 02138, USA \\
$^{2}$Department of Astronomy and Astrophysics, University of California, Santa Cruz, 1156 High Street, Santa Cruz, CA 95064 USA
}\date{December 2018}

\begin{document}

\maketitle

\label{firstpage}
\pagerange{\pageref{firstpage}--\pageref{lastpage}}

\begin{abstract} 
In this paper, we investigate whether galaxy assembly bias can reconcile the 20-40$\%$ disagreement between the observed galaxy projected clustering signal and the galaxy-galaxy lensing signal in the BOSS CMASS galaxy sample. We use the suite of \textsc{AbacusCosmos} $\Lambda$CDM simulations at Planck best-fit cosmology and two flexible implementations of extended halo occupation distribution (HOD) models that incorporate galaxy assembly bias to build forward models and produce joint fits of the observed galaxy clustering signal and the galaxy-galaxy lensing signal. We find that our models using the standard HODs without any assembly bias generalizations continue to show a 20-40$\%$ over-prediction of the observed galaxy-galaxy lensing signal. We find that our implementations of galaxy assembly bias do not reconcile the two measurements at Planck best-fit cosmology. In fact, despite incorporating galaxy assembly bias, the satellite distribution parameter, and the satellite velocity bias parameter into our extended HOD model, our fits still strongly suggest a $\sim34\%$ discrepancy between the observed projected clustering and galaxy-galaxy lensing measurements. 
It remains to be seen whether a combination of other galaxy
assembly bias models, alternative cosmological parameters, or baryonic
effects can explain the amplitude difference between the two signals. 

\end{abstract}
% so we need to write about the construction of the emulator and the tests

% do we want to talk about the failed trials?

\section{Introduction}

Weak gravitational lensing refers to the distortions in the images of distant galaxies by intervening mass along the line of sight. Because it directly measures the total mass distribution of the Universe, it has long been considered a powerful yet unique cosmological probe. Galaxy-galaxy lensing (hereafter ``g-g lensing") refers to the cross-correlation between foreground lens galaxy positions and the lensing shear of background source galaxies \citep{1984Tyson, 1996Brainerd, 1996dellAntonio, 2018Prat}. At small scales, it provides a measure of the radial distribution of total mass around galaxies, presenting an unique opportunity to directly probe the properties of dark matter halos. 

Recent surveys such as the Sloan Digital Sky Survey \citep[SDSS,][]{2000York}, Dark Energy Survey \citep[DES,][]{2016DES}, the Canada-France-Hawaii Telescope Lensing Survey \citep[CFHTlenS,][]{2012Heymans, 2013Erben}, the Kilo-Degree Survey \citep[KiDS,][]{2013deJong, 2015Kuijken}, and the Hyper Suprime Cam survey \citep[HSC,][]{2018Aihara} have generated thousands of square degrees of high signal-to-noise g-g lensing data \citep[e.g.][]{2013Mandelbaum, 2014Velander, 2018Abbott, 2018vanUitert}. Upcoming missions such as \textit{Euclid} \citep{2011Laureijs}, the Wide Field Infrared Survey Telescope \citep[WFIRST,][]{2013Spergel}, and the Large Synoptic Survey Telescope \citep[LSST,][]{2009LSSTL} promise to bring in even higher precision data over a vast fraction of the sky.

% now talk about wp data lalala
In parallel to these efforts, surveys such as the Baryon Oscillation Spectroscopic Survey \citep[BOSS,][]{2011Eisenstein, 2013Dawson}, have collected optical spectra for more than one million massive galaxies at $z < 1$. These spectra enabled accurate measurements of galaxy clustering in the form of the 2-point correlation function (2PCF), placing tight cosmology constraints. Upcoming experiments such as the Dark Energy Spectroscopic Instrument \citep[DESI,][]{2013Levi}, the Prime Focus Spectrograph \citep[PFS,][]{2014Takada}, and Euclid \citep{2011Laureijs} will measure the redshifts of tens of millions of galaxies, yielding exquisite measurements of galaxy clustering and also providing excellent lens samples for g-g lensing studies. 

While galaxy clustering and g-g lensing represent two independent yet complementary cosmology probes, \citet{2017Leauthaud} find discrepancies of 20-40 percent between their measurements of g-g lensing for CMASS galaxies and a model predicted from mock galaxy catalogs generated at Planck cosmology that match the CMASS projected correlation function \citep{2014Reid, 2016Saito}. \citet{2019Lange} extended this result by finding a similar $\sim 25\%$ discrepancy between the projected clustering measurement and the g-g lensing measurement in the BOSS LOWZ sample. They also found that this discrepancy is independent of redshift ($0.1 < z < 0.7$ and stellar mass ($11 < \log M_\star/M_\odot < 12$) in the BOSS CMASS and LOWZ sample. 
This discrepancy is well above the statistical error of the lensing signal and calls for a detailed re-examination of the forward model used to predict the g-g lensing signal. 

\citet{2017Leauthaud} found that lowering the cosmological parameter $S_8 = \sigma_8\sqrt{\Omega_m/0.3}$ by 2-3 $\sigma$ from the Planck 2015 value can reconcile the difference. However, cosmological effects are entangled with other effects due to details of galaxy-halo connection, baryons, and massive neutrinos. Before one can draw an inference about cosmological models, one must control for uncertainties in the astrophysical modeling of these other effects. 
%Baryon physics processes are a known source of uncertainties in these models. 
%BOSS CMASS mocks are generated using gravity-only N-body simulations, which do not account for any possible non-gravitational Baryonic effects. However, Baryon processes can affect the halo mass profile and the properties of sub-halos \citep[e.g.][]{2014vanDaalen, 2014Velliscig, 2016Chaves}. In fact, by comparing g-g lensing signals from the full-physics Illustris simulations \citep{2014aVogelsberger, 2014bVogelsberger, 2014Genel, 2015Sijacki, 2015Nelson} and the corrresponding gravity-only simulations, \citet{2017Leauthaud} found that Baryonic effects can induce a 10-30$\%$ effect on $\Delta\Sigma$, but the commonly used implementations of the halo-galaxy connection models are not yet flexible enough to account for these effects. 
One source of modeling uncertainties is the assumed connection between the observed galaxies
and their dark matter halos. \citet{2017Leauthaud} use a standard Halo Occupation Distribution (HOD) model of the \citet{2007Zheng} form. The mis-match in the amplitude of the g-g lensing signal may point to the failures of such empirical models. 

% assembly bias
In particular, one important aspect that these models neglect is galaxy assembly bias: the fact that in addition to halo mass, galaxy occupation depends on other properties such as halo age, spin, and concentration \citep[e.g.][]{2005Gao, 2006Wechsler, 2007Gao, 2007Zentner, 2008bDalal, 2011Lacerna, 2014Zentner}. Numerous recent studies have attempted using clustering data and simulations to detect galaxy assembly bias and constrain its effects \citep[e.g.][]{2016More, 2016Miyatake, 2016Saito, 2017Lehmann, 2018Xu, 2018Zehavi, 2019Contreras, 2019Zentner}. Galaxy assembly bias is especially relevant for the g-g lensing discrepancy because the clustering measurements tightly constrain the large-scale galaxy bias, whereas the lensing measurement is mostly sensitive to the dark matter halo mass profile. Thus, the g-g lensing discrepancy is fundamentally a discrepancy between halo mass and large-scale galaxy bias, the signature of assembly bias. 

In this paper, we explore the possibility of reconciling the g-g lensing discrepancy with two different extended HOD models incorporating galaxy assembly bias plus other halo scale physics. 
Specifically, we apply the generalized HOD model \citep[GRAND-HOD][]{2018Yuan}, which incorporates a novel implementation of galaxy assembly bias plus other generalizations to the \citet{2007Zheng} HOD, and the decorated HOD model \citep{2016Hearin}, which incorporates both a central assembly bias and a satellite assembly bias.
We build emulator models of the projected galaxy correlation function and the g-g lensing as a function of the extended HODs at Planck cosmology. We present joint fits to the observed projected galaxy correlation function and g-g lensing to evaluate how well the generalized HOD model can reconcile the two measurements. 

This paper is organized as follows. In Section~\ref{sec:theory}, we introduce the standard 5-parameter HOD, the generalized HOD, and the decorated HOD. In Section~\ref{sec:simNdata}, we present the galaxy clustering and weak lensing observables that we fit our extended HOD models to, and in Section~\ref{sec:emulator} we present our forward model for emulating the observables as a function of the extended HOD parameters. Then we use these emulator models to fit the observables and present the results in Section~\ref{sec:results}. We discuss the limitations and implications of our results in Section~\ref{sec:discussions}. Finally, in Section~\ref{sec:conclusions}, we summarize the main conclusions of our analysis.

Throughout this paper, unless otherwise specified, we assume a Planck 2015 \citep{2016Planck} cosmology with $H_0 = 67.26$~km/s/Mpc, $\Omega_\textrm{m} = 0.3141$, and $\sigma_{\textrm{8}} = 0.83$. We use halo mass definition $M_{200\textrm{b}}$, which we simply quote as $M$ for the rest of this paper.  

\section{Theoretical Background}
\label{sec:theory}

The standard 5-parameter HOD model \citep{2007Zheng} is a popular empirical framework to populate dark matter halos with mock central and satellite galaxies as a function of halo mass. However, for cosmology, this model may also be a source of systematics. In this section, we briefly review the standard HOD formalism and discuss physically motivated extensions to the standard HOD. 

\subsection{The standard HOD}

The standard HOD \citep{2007Zheng} gives the mean number of central galaxies and satellite galaxies as a function of halo mass
\begin{align}
&\bar{n}_{\mathrm{cent}} = \frac{1}{2}\mathrm{erfc} \left[\frac{\ln(M_{\mathrm{cut}}/M)}{\sqrt{2}\sigma}\right], \nonumber \\
& \bar{n}_{\textrm{sat}} = \left[\frac{M-\kappa M_{\textrm{cut}}}{M_1}\right]^{\alpha}\bar{n}_{\mathrm{cent}},
\label{equ:standard_hod}
\end{align}

where halo mass $M$ is again defined as $M_{200\textrm{b}}$.
The five parameters of this model are $M_{\textrm{cut}}, M_1, \sigma, \alpha, \kappa$. However, for the rest of the paper, we only consider the first four parameters of the model as our tests show that the predicted clustering and lensing observables depend very weakly on $\kappa$. The actual number of central galaxies in a halo follows the Bernoulli distribution. The actual number of satellites follows the Poisson distribution with the mean equal to $\bar{n}_{\textrm{sat}}$. The central assumes the location and velocity of the center-of-mass of the halo. The satellites are assigned to halo particles to track the dark matter distribution, with each particle of the halo having an equal probability of hosting a satellite galaxy.

\subsection{Generalized HOD}

To generate predictions for galaxy clustering, we populate
dark matter halos with mock galaxies using the publicly available \textsc{GRAND-HOD} package\footnote{\url{https://github.com/SandyYuan/GRAND-HOD}}. The routine introduces five new parameters to the standard HOD, including a novel implementation of galaxy assembly bias. \citet{2018Yuan} describes the generalizations in detail. We highlight the three parameters relevant for this study.

We introduce the satellite distribution parameter $s$, which deviates the satellite spatial distribution away from the halo profile. In our standard HOD implementation, satellites are placed on halo particles with equal probabilities. However, a positive $s$ leads the algorithm to favor particles further from halo center, effectively decreasing the concentration of satellites. Figure~2 of \citet{2018Yuan} shows how $s$ affects the predicted 2PCF. The range of $s$ is defined to be between $-1$ and 1. 

Similarly, we introduce the satellite velocity bias parameter $s_v$, which biases the satellite velocity distribution away from that of the halo. In our implementation, the satellites always assume the velocities of the particles they are placed on. A positive $s_v$ simply favors particles with higher velocity relative to the halo center to host satellite galaxies. By making sure that each galaxy still tracks the velocity and position of a dark matter particle, this implementation guarantees that the satellite galaxies still obey Newtonian physics in the halo potential. 
While peculiar velocities do not directly affect the projected correlation function, our implementation of velocity bias does change the clustering because particle velocity relative to halo center is correlated to particle position in the halo, and the higher velocity subsample of particles do not evenly trace the density profile of the halo. Figure~4 of \citet{2018Yuan} shows how $s_v$ affects the predicted 2PCF. The range of $s_v$ is defined to be between $-1$ and 1. 

We also introduce a galaxy assembly bias parameter $A$. We set the NFW \citep{1997Navarro} halo concentration as the secondary dependence for the galaxy occupation besides halo mass. The NFW concentration is defined as 
\begin{equation}
    c = \frac{r_{\textrm{vir}}}{r_{s, \textrm{Klypin}}},
    \label{equ:concentration}
\end{equation}
where $r_{\textrm{vir}}$ is the virial radius of the halo and $r_{s, \textrm{Klypin}}$ is the Klypin scale radius \citep{2011Klypin}.
In our implementation, we first rank all halos by halo mass and calculate the number of galaxies $n_{\textrm{gal}}$ for each halo. We save the list of $n_{\textrm{gal}}$. Then we re-rank the halos according to a ``pseudo-mass" defined as 
  \begin{equation}
  \log M_{\textrm{pseudo}} = \begin{cases} 
  \log M + A & \text{if}\ \  c > \bar{c}(M), \\
   \log M - A & \text{if}\ \ c < \bar{c}(M),
   \end{cases}
  \label{equ:Mpseudo}
  \end{equation}
  
where $A$ is the assembly bias parameter which governs the strength of assembly bias in our model, and $\bar{c}(M)$ is the median concentration within a mass bin at mass $M$. Note that we do not re-rank the $n_{\textrm{gal}}$ list, just the halos themselves. Finally, we assign the numbers in the $n_{\textrm{gal}}$ list to the re-ranked list of halos in order. Effectively, for a positive $A$, we are swapping galaxies in a more massive less concentrated halo to a less massive more concentrated halo. This swapping routine ensures that the total number of galaxies is preserved in the catalog when only the assembly bias parameter is varied. However, it does not preserve the expected number of galaxies for a given halo mass $\langle \bar{n}_g|M\rangle$, in contrast to the standard assembly bias interpretation. 
Figure~6 of \citet{2018Yuan} shows the effect of $A$ on the predicted 2PCF. The range of $A$ is technically between $-\infty$ and $\infty$, but we expect $A$ to be on the order of $10^{-1}$. 

In this paper we use a generalized HOD with 7 parameters: $M_{\textrm{cut}}$, $M_1$, $\sigma$, $\alpha$, $s$, $s_v$, and $A$. Again, we have ignored parameter $\kappa$. Our goal is to emulate the projected 2PCF and the g-g lensing signal as a function of these HOD parameters to search for good fits to both measurements within this generalized model space. 

Again we stress that for this study we use $M_{200b}$ as our halo mass, whereas the publicly available GRAND-HOD code uses the virial mass $M_{vir}$. We also use a slightly modified formula for the mean number of satellites $\bar{n}_{\textrm{sat}}$ compared to that available on GRAND-HOD. Namely the GRAND-HOD uses 
\begin{equation}
\bar{n}_{\textrm{sat}} = \left[ \frac{M - \kappa M_{\textrm{cut}}}{M_1}\right]^\alpha,
\end{equation}
whereas for this study we modulate the number of satellites with the number of centrals $\bar{n}_{\textrm{cent}}$ (see Equation~\ref{equ:standard_hod}).
We make these adjustments to be more consistent with the HOD implementation of \citet{2017bAlam}, which fitted the standard HOD model to the CMASS projected 2PCF, though our implementation still differs in key aspects such as the satellite profile.

\subsection{Decorated HOD}

The decorated HOD \citep[][]{2016Hearin} provides another way of incorporating assembly bias into the standard HOD. The decorated HOD is conveniently implemented in the \texttt{Halotools} code package \citep{2017Hearin}. We borrow a slightly modified implementation of the decorated HOD implementation from \citet{2019Wang}, which also uses the NFW halo concentration (Equation~\ref{equ:concentration}) as the secondary property of the halo. 
Again, they assume that $P(n_{\textrm{cen}}|M, c)$
is a Bernoulli distribution and that $P(n_{\textrm{sat}}|M, c)$ is a Poisson distribution, but that these distributions have first moments of
\begin{align}
    \langle n_{\textrm{gal}} | M, c > c_{\textrm{piv}}\rangle =& \langle n_{\textrm{gal}} | M \rangle + \delta n_{\textrm{gal}} \nonumber\\
    \langle n_{\textrm{gal}} | M, c < c_{\textrm{piv}}\rangle =& \langle n_{\textrm{gal}} | M \rangle - \delta n_{\textrm{gal}}.
\end{align}
where the $n_{\textrm{gal}}$ notation applies to both the centrals and satellites. The pivotal value $c_{\textrm{piv}}$ is chosen to be the median concentration at a given halo mass. This implementation can be conceptualized as the top $50\%$ of halos in concentration in a mass bin takes some galaxies away from the bottom $50\%$. The magnitude of $\delta n_{\textrm{cent}}$ and $\delta n_{\textrm{sat}}$ are characterized by 
\begin{align}
    &\delta n_{\textrm{cent}} =  A_{\textrm{cent}} \textrm{min} \left[ \langle n_{\textrm{cent}}|M\rangle, 1 - \langle n_{\textrm{cent}}|M\rangle\right] \nonumber \\
    &\delta n_{\textrm{sat}} = A_{\textrm{sat}} \langle n_{\textrm{sat}}| M\rangle
\end{align}
where $A_{\textrm{cent}}$ and $A_{\textrm{sat}}$ are the central assembly bias and satellite assembly bias parameters, respectively.
The two assembly bias parameters both range between $-1$ and 1. This decorated HOD implementation also introduces modifications to the standard \texttt{Halotools} implementation, in that it preserves the total number of galaxies in the mock catalogs that differ only in their assembly bias parameter values. This is achieved by conditioning the decorated HOD on the total number of galaxies before populating each individual halos with galaxies. This modification shifts the number of centrals from a Bernoulli distribution but preserves the Poisson distribution in the number of satellites in each halo. 

\section{Simulations and Data}
\label{sec:simNdata}
In this section, we introduce the galaxy projected clustering and weak lensing observables that we fit our different HOD models to. We also discuss the simulations used to predict the observables.

\subsection{Galaxy clustering and g-g lensing data}

The first observable we consider in this paper is the projected galaxy 2PCF, commonly referred to as $w_p$. It is defined as 
\begin{equation}
w_p(r_\perp) = 2\int_0^{\pi_{\textrm{max}}} \xi(r_\perp, \pi)d\pi,
\label{equ:wp_def}
\end{equation}
where $\xi(r_\perp, \pi)$ is the anisotropic 2PCF, and $r_\perp$ and $\pi$ are transverse and line-of-sight (LOS) separations in comoving units. In this paper, we will be matching our theory $w_p$ to the observed $w_p$ by \citet{2016Saito}, which was measured on the BOSS DR10 CMASS sample in the redshift range $0.43 < z < 0.7$. The associated covariance matrix was determined by \citet{2014Reid}. The covariance matrix is computed from 5,000,000 realizations drawn from 200 bootstrap regions in the survey of roughly equal size and shape. 
Note that the $w_p$ measurement of \citet{2016Saito} assumes a different fiducial cosmology with $H_0 = 70$~km/s/Mpc and $\Omega_{\textrm{m}} = 0.274$. Our models generate the predicted $w_p$ in Planck 2015 cosmology but we convert the prediction to the \citet{2016Saito} cosmology using a set of simple conversion formulas presented in \citet{2013bMore}. Our $w_p$ plots throughout this paper are shown assuming the \citet{2016Saito} fiducial cosmology.

%\begin{figure}
%    \centering
%    \hspace*{-0.7cm}
%    \includegraphics[width=3.8in]{./plot_cov_rwp.pdf}
%    \vspace{-0.7cm}
%    \caption{The correlation matrix of the projected correlation function $w_p$. The small-scale region is shot noise dominated and thus shows little correlation whereas the large-scale region is heavily correlated due to sample variance. }
%    \label{fig:cov_wp}
%\end{figure}

%\begin{figure}
%    \centering
%    \hspace*{-0.7cm}
%    \includegraphics[width=3.8in]{./plot_cov_rdelsig.pdf}
%    \vspace{-0.7cm}
%    \caption{The correlation matrix of the g-g lensing measurement %$\Delta\Sigma$. The large-scale bins are moderately correlated due to sample variance and correlated shape noise. The small-scale bins are uncorrelated where the errors are dominated by shape noise. \red{Sandy: yeah we could take out these cov matrix plots. Daniel, what do you think?}}
%    \label{fig:cov_ds}
%\end{figure}

The g-g lensing observable we use is the mean surface mass density contrast profile $\Delta \Sigma$, defined as
\begin{equation}
    \Delta\Sigma(r_\perp) = \overline{\Sigma}(<r_\perp) - \overline{\Sigma}(r_\perp),
\label{equ:ds_def}
\end{equation}
where $\overline{\Sigma}(r_\perp)$ is the azimuthally averaged and projected surface mass density at radius $r_\perp$ and $ \overline{\Sigma}(<r_\perp)$ is the mean projected surface mass density within radius $r_\perp$ \citep{1991Miralda-Escude, 2001Wilson, 2017Leauthaud}. The observed $\Delta\Sigma$ signal is presented in \citet{2017Leauthaud} for a flat $\Lambda$CDM cosmology with $H_0 = 100$~km/s/Mpc and $\Omega_{\textrm{m}} = 0.31$, where the lens galaxy sample is the BOSS DR10 CMASS sample between $0.43 < z < 0.7$ and the background galaxy sample is a combination of two datasets: the Canada France Hawaii Telescope Lensing Survey \citep[CFHTLenS,][]{2012Heymans, 2013Miller} and the Canada France Hawaii Telescope Stripe 82 Survey \citep[CS82,][Erben et al. in prep]{2017Leauthaud}. 
The covariance matrix, computed via bootstrap, is presented in \citet{2017Leauthaud}. 

\subsection{Simulation and Mocks}
\label{subsec:mocks}

For the purpose of this paper, we use a series of galaxy mocks generated from the \textsc{AbacusCosmos} N-body simulation suite, generated by the fast and high-precision \textsc{Abacus} N-body code \citep[][Ferrer et al., in preparation; Metchnik $\&$ Pinto, in preparation]{2018Garrison, 2016Garrison}. We use 20 boxes of comoving size 1100~$h^{-1}$Mpc with Planck 2015 cosmology \citep{2016Planck} at redshift $z = 0.5$. These boxes are set to different initial phases to generate unique outputs. Each box contains 1440$^3$ dark matter particles of mass $4\times 10^{10}$ $h^{-1}M_{\odot}$. The force softening length is 0.06~$h^{-1}$~Mpc. Dark matter halos are found and characterized using the ROCKSTAR \citep{2013Behroozi} halo finder. 

\begin{table}
\centering
\begin{tabular}{ c | c }
\hhline {==}
Parameter name & Baseline value \\ 
\hline
$\log_{10}(M_{\textrm{cut}}/h^{-1}M_\odot)$ & $13.248$  \\ 
$\log_{10}(M_1/h^{-1}M_\odot)$ & 14.179 \\
$\sigma$ & 0.897 \\
$\alpha$ & 1.151 \\
$\kappa$ &  0.137 \\
\hline 
\end{tabular} 

\caption{The standard HOD parameters and their baseline values as quoted from \citet{2017bAlam}.}
\label{tab:baseline}
\end{table}

To generate mock galaxies, we implement 421 generalized HODs: 1 baseline HOD \citep[the values of the 5 standard parameters are taken from][as shown in Table~\ref{tab:baseline}, with the generalized parameters set to 0]{2017bAlam} and 210 pairs of perturbed HODs. 
Each perturbed pair consists of two generalized HODs symmetrically perturbed around the baseline values. The perturbation is uniformly sampled from a high-dimensional box in the generalized HOD parameter space, hard bounded in each parameter direction. The 210 pairs of perturbed HOD are divided into three tiers, with increasing bounds in parameter space. Tier 1 consists of the first 15 HOD pairs and sample within the tightest bounds around the baseline HOD in parameter space; tier 2 consists of the next 75 pairs of HODs and sample from a larger box in parameter space; tier 3 consists of the final 120 pairs of HODs and sample from the largest box in parameter space. This tiered approach guarantees a higher density of samples in the region closest to the baseline value, effectively increasing the weight of the closest region in the fits. Table~\ref{tab:hodranges} lists the specific bounds in each parameter for each tier.

\begin{table}
\centering
\begin{tabular}{ c | c | c | c}
\hhline {====}
Parameter & Tier 1 (15 pairs) & Tier 2 (75 pairs) & Tier 3 (120 pairs)\\ 
\hline
$\log_{10}M_{\textrm{cut}}$ & [13.225, 13.269] & [13.151, 13.327] & [12.748, 13.748]  \\ 
$\log_{10}M_1$ & [14.157, 14.200] & [14.082, 14.258] & [13.679, 14.679] \\
$\sigma$ & [0.847, 0.947] & [0.697, 1.097] & [0.497, 1.297] \\
$\alpha$ & [1.101, 1.201] & [0.951, 1.351] & [0.751, 1.551] \\
$s$ &  [-0.05, 0.05] & [-0.2, 0.2] & [-0.3, 0.3] \\
$A$ &  [-0.05, 0.05] & [-0.2, 0.2] & [-0.3, 0.3] \\
\hline 
\end{tabular} 

\caption{The parameter ranges for the 3 tiers of HOD pairs we have generated for training the generalized HOD emulator. Each bracket shows the minimum and maximum bounds in that parameter for that tier. Our mocks are generated from HODs uniformly sampled from within these bounds. As discussed in the main text, each tier increases the bounds in all the HOD parameters, and this tiered approach ensures a higher density of samples in the region closest to the baseline HOD. The $\log_{10}M_{\textrm{cut}}$ notation is a shorthand for $\log_{10}(M_{\textrm{cut}}/h^{-1}M_\odot)$ for brevity.}
\label{tab:hodranges}
\end{table}

Each HOD is run over 20 simulation boxes and in each box repeated 4 times with 4 different random number generator seeds. We take the average over the 4 runs with different seeds and 20 boxes to reduce sample variance. 

To generate mock projected correlation function $w_p$, we run \textsc{Corrfunc} \citep{manodeep2016} with $\pi_{\textrm{max}} = 77.6 h^{-1}$~Mpc (this value in Planck cosmology matches $\pi_{\textrm{max}} = 80h^{-1}$~Mpc used in \citet{2016Saito} in their fiducial cosmology) on the mock galaxies to obtain the $w_p$. 
We use 18 evenly spaced logarithmic bins in $r_\perp$ between $r_\perp = 0.165 h^{-1}$~Mpc and $r_\perp = 29.3 h^{-1}$~Mpc. These values are again chosen to match those of \citet{2016Saito}. 
To compute the mock g-g lensing signal, we use the $\Delta \Sigma$ functionality provided in \texttt{Halotools}, with 10 logarithmically spaced bins between $r_\perp = 0.157 h^{-1}$~Mpc and $r_\perp = 15 h^{-1}$~Mpc, matched with those used \citet{2017Leauthaud}. We do not go below $0.157 h^{-1}$~Mpc due to limited force softening resolution in our simulations.

Before we move on to the emulator fits, we perform a rejection step for our samples of paired HODs. Specifically, having computed the $w_p$ of each of the 421 HODs, we reject the ones whose $w_p$ deviate too far from the baseline $w_p$. The reason for this rejection step is that there are directions in the HOD space that drastically change the $w_p$, way beyond the capabilities of our emulator model (described in the following section). Even within our HOD parameter sample, we find examples where a combination of high $M_\mathrm{cut}$ and low $M_1$ leads to a very high satellite fraction, resulting in a $w_p$ that is 10 times higher in amplitude than our baseline $w_p$. To constrain ourselves to regions of the HOD space with a moderately behaved $w_p$, we choose an arbitrary but generous cut of 
\begin{equation}
    -0.2 < \log_{10}\left(\frac{w_p(r_\perp<1 h^{-1}\textrm{Mpc})}{w_{p,0}(r_\perp<1 h^{-1}\textrm{Mpc})}\right)< 0.2,
\label{equ:wpcut}
\end{equation}
where $w_p(r_\perp<1 h^{-1}\textrm{Mpc})$ is the $w_p$ integrated up to $r_\perp = 1 h^{-1}\textrm{Mpc}$, and $w_{p,0}$ denotes the baseline $w_p$. This cut roughly corresponds to a percentage range of $63\% - 158\%$ of the baseline $w_p$, more than sufficient to cover the a few percent level uncertainties in the observed $w_p$. By applying this cut, we remove 155 HODs from our sample of mocks, with the remaining 266 HODs moving on to the emulator fits. 

To generate mocks for the decorated HOD emulator, we follow the same procedure. We first generate 1 baseline HOD (using the HOD parameters taken from \citet{2017bAlam} with the assembly bias parameters set to 0), and 240 pairs of perturbed HODs around the baseline. The perturbed HODs are divided into four tiers, with each HOD applied to all 20 simulation boxes with 4 random number seeds. We summarize the parameter ranges in Table~\ref{tab:hodranges_dec}.

\begin{table*}
\centering
\begin{tabular}{ c | c | c | c | c }
\hhline {=====}
Parameter & Tier 1 (15 pairs) & Tier 2 (45 pairs) & Tier 3 (120 pairs) & Tier 4 (120 pairs)\\ 
\hline
$\log_{10}M_{\textrm{cut}}$ & [13.225, 13.269] & [13.151, 13.327] & [13.048, 13.448] & [12.748, 13.748]  \\ 
$\log_{10}M_1$ & [14.157, 14.200] & [14.082, 14.258] & [13.979, 14.379] & [13.679, 14.679] \\
$\sigma$ & [0.847, 0.947] & [0.697, 1.097] & [0.697, 1.097] & [0.497, 1.297] \\
$\alpha$ & [1.101, 1.201] & [0.951, 1.351] & [0.951, 1.351] & [0.751, 1.551] \\
$A_\textrm{cent}$ &  [-0.05, 0.05] & [-0.2, 0.2] & [-0.2, 0.2] & [-0.3, 0.3]\\
$A_\textrm{sat}$ &  [-0.05, 0.05] & [-0.2, 0.2] & [-0.2, 0.2] & [-0.3, 0.3] \\
\hline 
\end{tabular} 
\caption{The parameter ranges for the 4 tiers of HOD pairs we have generated for training the decorated HOD emulator. Each bracket shows the minimum and maximum bounds in that parameter for that tier. Our mocks are generated from HODs uniformly sampled from within these bounds. The $\log_{10}M_{\textrm{cut}}$ notation is a shorthand for $\log_{10}(M_{\textrm{cut}}/h^{-1}M_\odot)$ for brevity.}
\label{tab:hodranges_dec}
\end{table*}

Then we perform the same selection step to select only the mocks whose $w_p$ is within a reasonable range of the baseline $w_p$. We employ the following cut
\begin{equation}
    -0.15 < \log_{10}\left(\frac{w_p(r_\perp<1 h^{-1}\textrm{Mpc})}{w_{p,0}(r_\perp<1 h^{-1}\textrm{Mpc})}\right)< 0.15,
\label{equ:wpcut_dec}
\end{equation}
 and we select our final training set of 251 HODs.

\section{Methods}
\label{sec:emulator}

In this section, we present our methodology for constructing the $w_p$ and $\Delta\Sigma$ emulators from simulations. Then we discuss the use of nested sampling to explore the generalized HOD parameter space when fitting the observed $w_p$ and $\Delta\Sigma$.

\subsection{The $w_p$ and $\Delta\Sigma$ emulator}
\label{subsec:model}
We first construct an emulator that models the galaxy projected 2PCF $w_p$ as a function of the generalized HOD parameters ($\log_{\textrm{10}} M_{\textrm{cut}}, \log_{\textrm{10}} M_1, \sigma, \alpha, s, s_v, A$). We model the weighted projected 2PCF $r_\perp w_p$ because it has more moderate behavior, resulting in a more balanced covariance matrix. 

There is a variety of ways to construct the model such as neural nets and gaussian processes. Here we are using a model based on first and second derivatives. Specifically, we can write down our model analytically as 
\begin{align}
\centering
r_\perp w_{p,i}(\mathbf{p}_0+\delta \mathbf{p}) = & r_\perp w_{p,i}(\mathbf{p}_0)
+ \Sigma_{j = 1}^{7} \frac{\partial r_\perp w_{p,i}}{\partial p_j} \delta p_j \nonumber\\ 
&+ \frac{1}{2}\Sigma_{j,k = 0}^{7} \frac{\partial^2 r_\perp w_{p,i}}{\partial p_j \partial p_k}\delta p_j\delta p_k, \label{equ:model_wp}
\end{align}
where $i$ is the bin number and $\mathbf{p} = [p_1, p_2, ..., p_7]$ are the 7 HOD parameters. Thus, for each bin, we have a total of 36 coefficients to fit for: the intercept $r_\perp w_{p,i}(\mathbf{p}_0)$, the 7 first derivatives $\partial r_\perp w_{p,i}/\partial p_j$, and the 28 second derivatives $\partial^2 r_\perp w_{p,i}/\partial p_j \partial p_k$. 

Similarly, we can construct our g-g lensing model with the following formula
\begin{align}
\centering
r_\perp \Delta\Sigma_{i}(\mathbf{p}_0+\delta \mathbf{p}) = & r_\perp \Delta\Sigma_{i}(\mathbf{p}_0)
+ \Sigma_{j = 1}^{7} \frac{\partial r_\perp\Delta\Sigma_{i}}{\partial p_j} \delta p_j \nonumber\\ 
&+ \frac{1}{2}\Sigma_{j,k = 0}^{7} \frac{\partial^2 r_\perp \Delta\Sigma_{i}}{\partial p_j \partial p_k}\delta p_j\delta p_k, \label{equ:model_delsig}
\end{align}
where we have again modulated the lensing signal $\Delta\Sigma$ with $r_\perp$ to produce a more flat behavior as a function of scale and a more balanced covariance matrix.

\begin{figure}
    \centering
    \hspace*{-1cm}
    \includegraphics[width=3.5in]{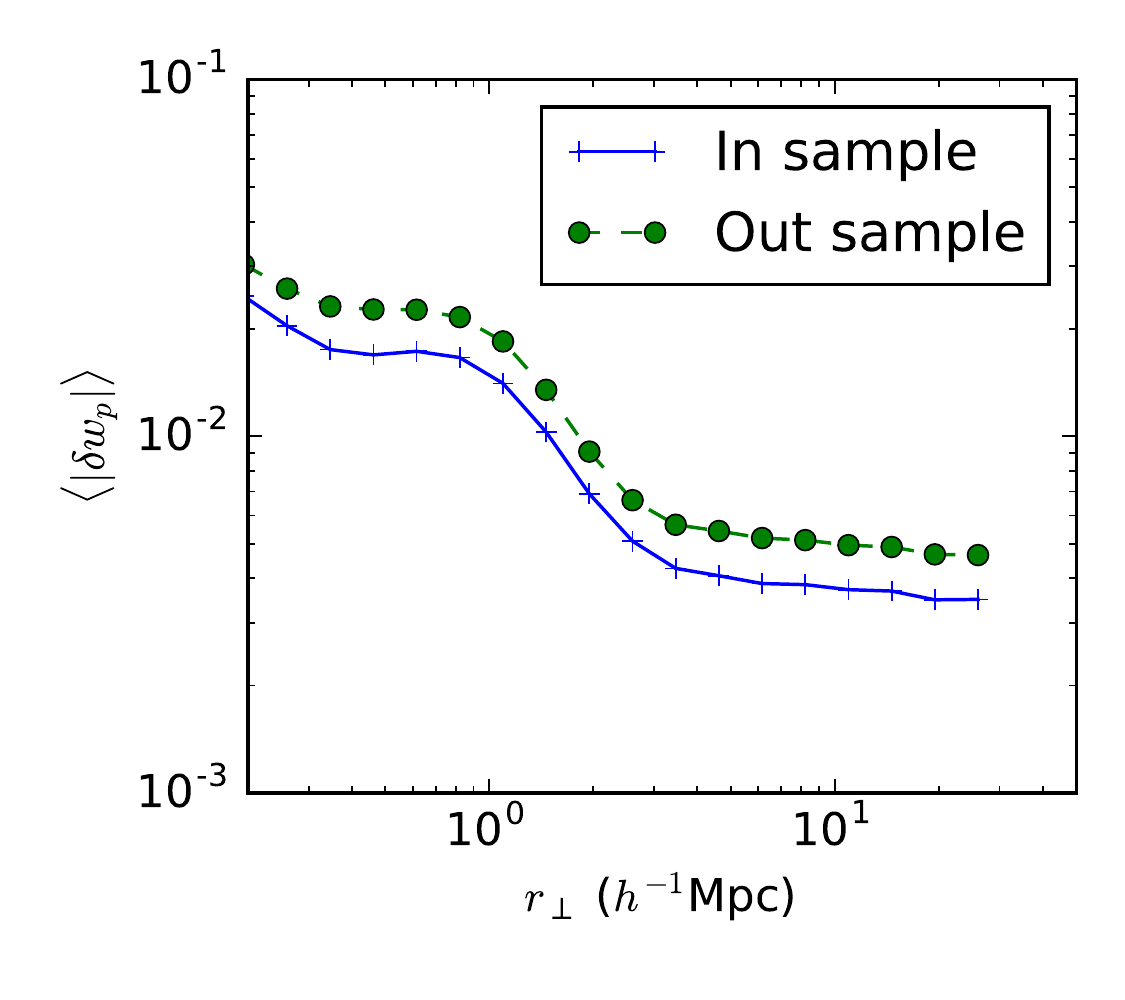}
    \vspace{-0.7cm}
    \caption{The absolute in-sample errors and out-sample errors of our best-fit $w_p$ emulator, averaged across all test HODs and across 10 validation runs. The $\delta$ notation denotes that the errors are in fraction of $w_p$ itself. We see the maximum out-sample error exists at small scales and is around $2.3\%$. The in-sample error is somewhat smaller than the out-sample error overall. }
    \label{fig:cv_wp}
\end{figure}

\begin{figure}
    \centering
    \hspace*{-1cm}
    \includegraphics[width=3.5in]{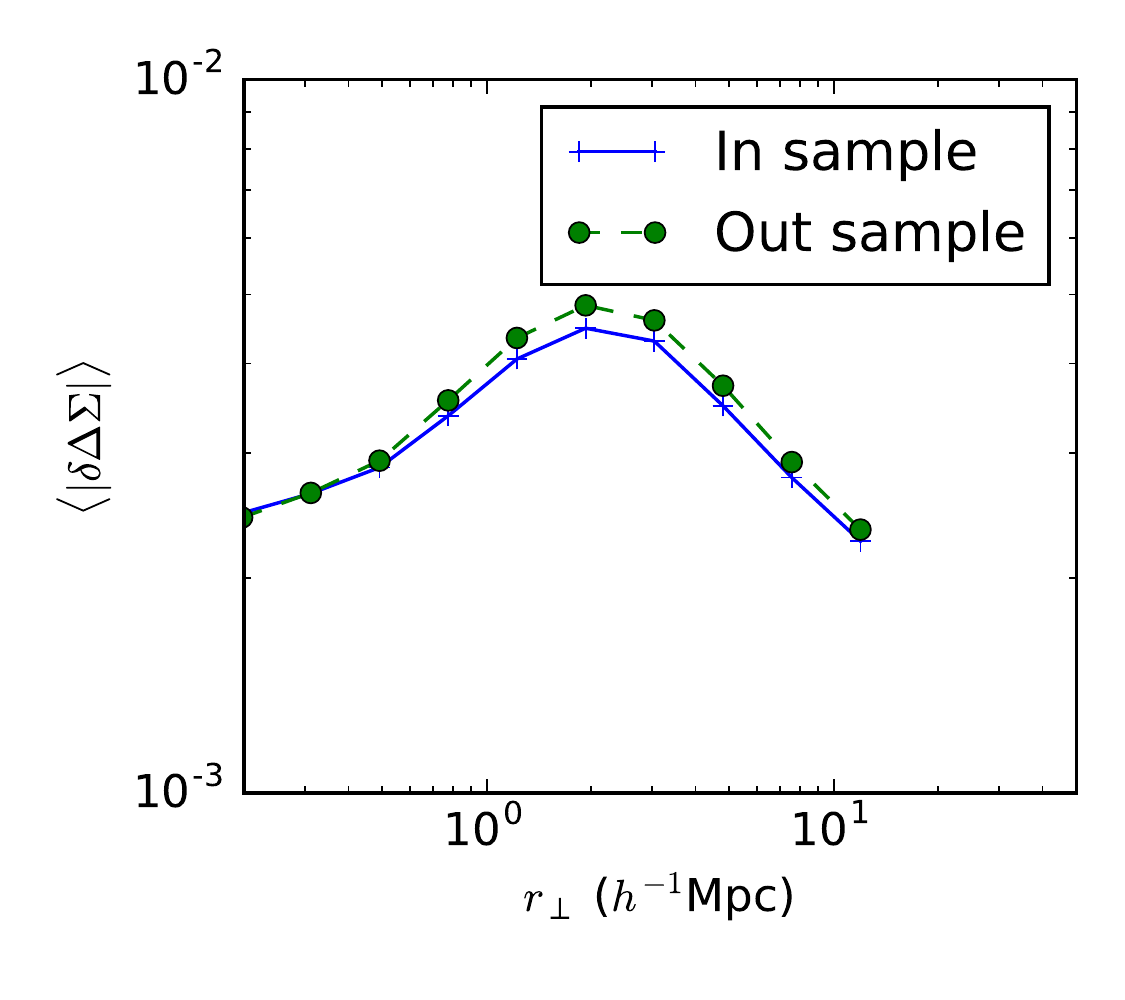}
    \vspace{-0.7cm}
    \caption{The absolute in-sample errors and out-sample errors of our best-fit $\Delta\Sigma$ emulator, averaged across all test HODs and across 10 validation runs. The $\delta$ notation denotes that the errors are in fraction of $\Delta\Sigma$ itself. We see the out-sample error is $< 0.5\%$ and does not seem to strongly depend on scale. The in-sample error is about equal to the out-sample error. }
    \label{fig:cv_ds}
\end{figure}

We fit the models given in Equation~\ref{equ:model_wp} and Equation~\ref{equ:model_delsig} to the mock observables presented in Section~\ref{subsec:mocks} using a standard least-squares routine. The two emulators are fitted individually and each $r_\perp$ bin is also fitted independently. For each emulator, each bin in $r_\perp$ has 36 unknowns and 266 data points in the fit. 

We run a set of cross validation tests to examine the performance of our fit. We perform 10 cross validation runs. For each run, we train our model on a randomly selected $\sim90\%$ of HODs plus the baseline HOD and then test on the remaining $\sim10\%$. This way, we can compute the in-sample and out-sample errors of our fit. The in-sample errors for each validation fun are computed by testing our best-fit emulator for that run against the set of generalized HODs that the model is trained on, while the out-sample errors are computed by testing against the set of generalized HODs that the model is not trained on.  

We show the average of the absolute values of the in-sample errors and out-sample errors of our best-fit emulators as a function of $r_\perp$ in Figure~\ref{fig:cv_wp} and Figure~\ref{fig:cv_ds}, respectively. The average is calculated by averaging the fractional error across all generalized HODs in the test set, then averaged over all 10 cross-validation runs. The errors shown are in fraction of $w_p$ itself.

To reduce overfitting in the $w_p$ emulator, we calculate the mean and standard deviation of the best-ft of each of the 36 coefficients by combining the best fit values from the 10 cross-validation runs. We set to zero the 2 coefficients that have the lowest signal noise. These 2 coefficients correspond to the following second derivatives: $\partial s_v \partial A$ and $\partial \log_{10}M_1 \partial s_v$. Dropping these 3 second derivatives gives us a slight reduction in out-sample error by $\approx 2\%$. The errors shown in Figure~\ref{fig:cv_wp} are the errors after removing these 2 second derivatives from the emulator. We do not repeat this procedure for the g-g lensing emulator as it has low out-sample errors and does not appear to be overfit. 

We repeat the same procedure for the decorated HOD to construct emulators of $w_p$ and $\Delta \Sigma$ as a function of 6 decorated HOD parameters ($\log_{\textrm{10}} M_{\textrm{cut}}, \log_{\textrm{10}} M_1, \sigma, \alpha, A_{\textrm{cent}}, A_{\textrm{sat}}$). We run cross-validation tests and remove second derivatives $\partial \log_{\textrm{10}} M_\textrm{cut} \partial A_{\textrm{sat}}$ from the $w_p$ emulator due to their low signal-to-noise. We recover a relative out-sample error of $< 2.1\%$ in the $w_p$ emulator and $< 0.5\%$ in the $\Delta \Sigma$ emulator. 

\subsection{The likelihood functions}
\label{subsec:likelihood}

The joint log-likelihood function is computed as 
\begin{equation}
    \log L(D|\Theta, M) = \log L(w_p|\Theta, M) + \log L(\Delta\Sigma|\Theta, M),
    \label{equ:jointL}
\end{equation}
where the likelihood functions for $w_p$ and $\Delta\Sigma$ are assumed to be Gaussian. Thus, we have 
\begin{align}
    \log L(w_p|\Theta, M) = - \frac{1}{2} \delta \mathbf{w}_p^T \mathbf{C}^{-1}\delta \mathbf{w}_p,
    \label{equ:logL_wp}
\end{align}
where $\mathbf{C}$ is the observed covariance matrix of $w_p$, and $\delta \mathbf{w}_p$ is the difference vector between the observed $w_p$ and the emulated $w_p$ given the HOD parameters,
\begin{equation}
    \delta \mathbf{w}_p = \delta \mathbf{w}_p^{\textrm{obs}} - \delta \mathbf{w}_p^{\textrm{emulated}}(\Theta).
\end{equation}
We construct the log-likelihood of $\Delta\Sigma$ in analogous manner.

\subsection{Nested Sampling}
\label{subsec:dynesty}

We need a sampling algorithm to explore the generalized HOD posterior space, which we compute from the likelihood function and the priors. In this paper, we would also like to compare different generalized HOD models using Bayesian evidence. The Bayesian evidence is defined as
\begin{equation}
    \mathcal{Z} = P(D|M) = \int_{\Omega_{\Theta}} P(D|\Theta, M)P(\Theta|M)d\Theta.
    \label{equ:evidence}
\end{equation}
where $M$ represents the model, $D$ represents the data, and $\Theta$ represents the model parameters. $P(D|\Theta, M)$ is the likelihood of the data given the parameters of our model, and $P(\Theta|M)$ is the prior for the parameters of our model. The evidence can simply be interpreted as the marginal likelihood of the data given the model, and serves as an important metric in Bayesian model comparisons. 

The nested sampling technique, first developed by \citet{2006Skilling} gives us an effective way to compute Bayesian evidence integrals. For this paper, we use the publicly available nested sampling code \texttt{dynesty} \citep{2018Speagle, 2019Speagle}. This code computes the Bayesian evidence while generating samples of the posterior parameter space. It replaces the multi-dimensional evidence integral over model parameters in Equation~\ref{equ:evidence} with a 1D integral over the prior mass contained within nested isolikelihood contours. For a detailed description of the code, refer to \citet{2019Speagle}.

In our \texttt{dynesty} runs, we use a nested sampler with 1500 live points and a random walk sampler conditioned on the bounding distribution. The stopping criterion is set to $d\log\mathcal{Z} > 0.01$.

\section{Results}
\label{sec:results}

In this section, we use the $w_p$ and $\Delta \Sigma$ emulators to simultaneously fit the observed $w_p$ and $\Delta \Sigma$ signals and to explore the generalized/decorated HOD parameter space, using the likelihood function described in Section~\ref{subsec:likelihood} and the nested sampling technique described in Section~\ref{subsec:dynesty}. 

In the following subsections, we first showcase the marginalized parameter constraints for the generalized HOD model, then the marginalized parameter constraints for the decorated HOD model. We compare the two models and discuss whether either of them provides a good simultaneous fit of the observed projected clustering and g-g lensing signal. 

\subsection{Joint fits with generalized HOD}
\label{subsec:fits_genHOD}

\begin{figure*}
    \centering
    \hspace*{-0.7cm}
    \includegraphics[width=7.6in]{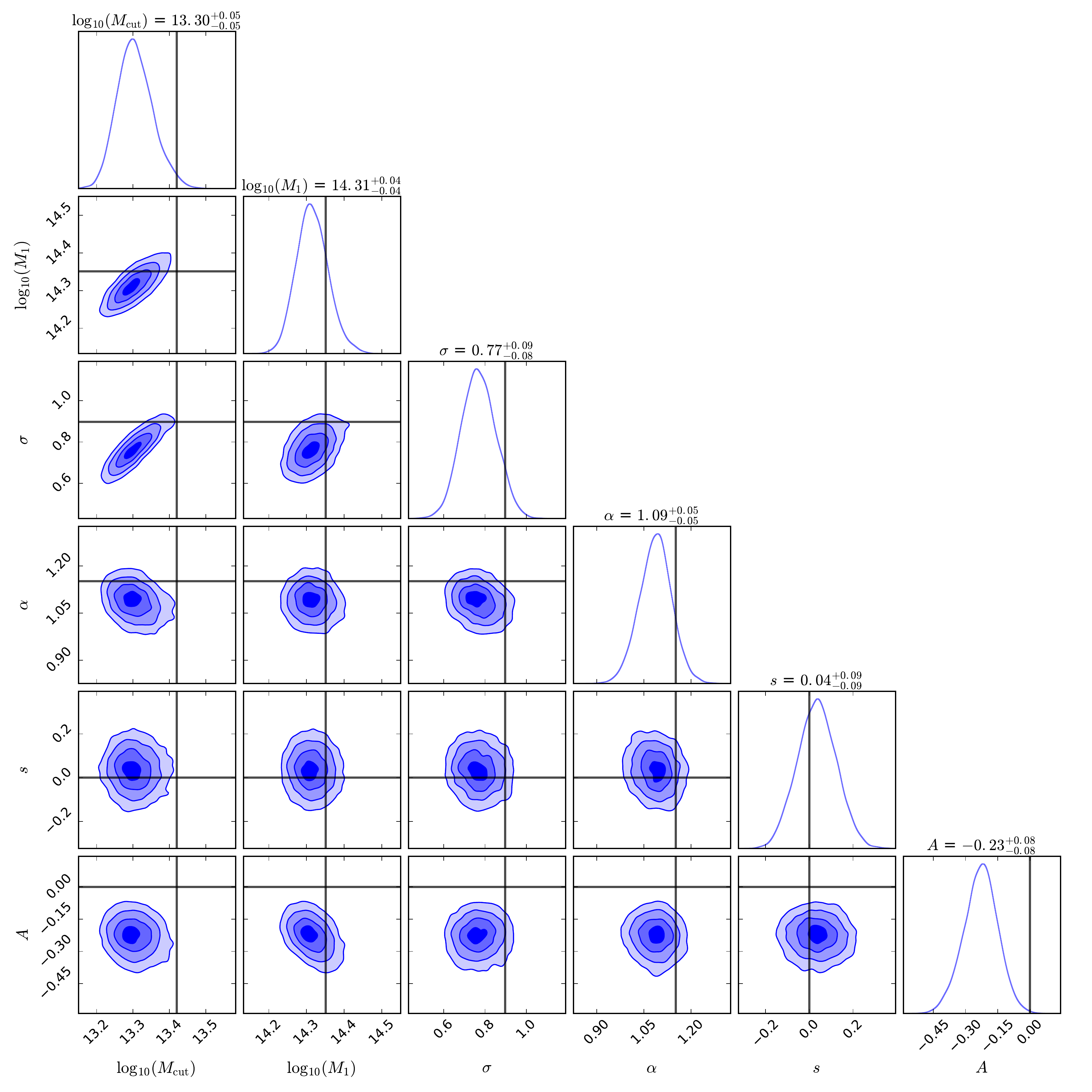}
    \vspace{-0.7cm}
    \caption{The 1D and 2D marginalized posterior constraints on the generalized HOD parameters. The contours shown correspond to 0.5, 1, 1.5, and 2 $\sigma$ uncertainties. The vertical and horizontal lines show the centers of the Gaussian priors for reference. The values displayed above the 1D marginals are posterior medians with the upper/lower bounds associated with the 0.025 and 0.975 quantiles.}
    \label{fig:corner_genhod}
\end{figure*}

\begin{figure*}
    \centering
    \hspace*{-0.7cm}
    \includegraphics[width=7in]{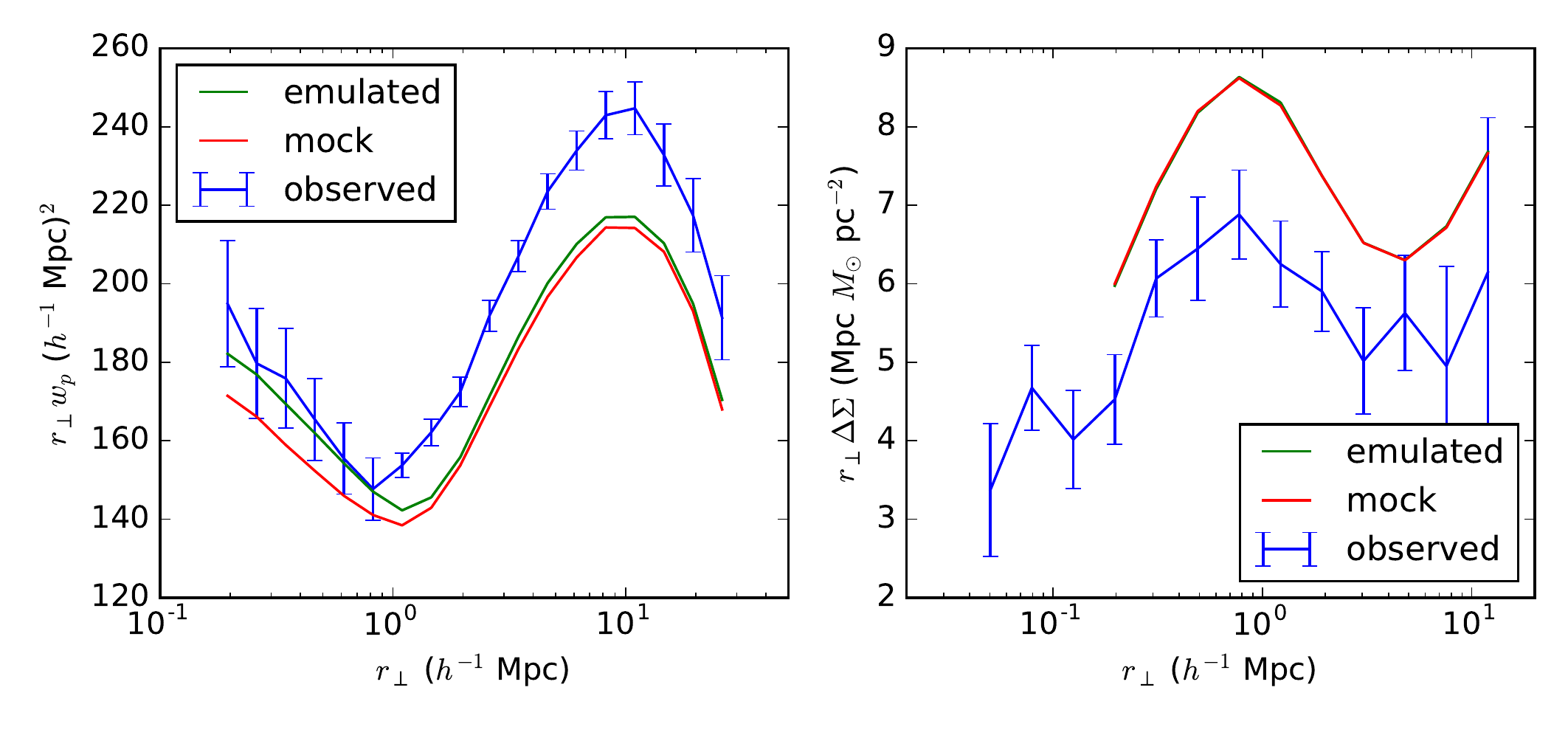}
    \vspace{-0.7cm}
    \caption{The maximum a posteriori (MAP) emulated $w_p$ and $\Delta \Sigma$ (in green) compared to the observed $w_p$ and $\Delta \Sigma$ (in blue) and the mock $w_p$ and $\Delta \Sigma$ (in red), using the 6-parameter generalized HOD model. The mock observables are computed from mock catalogs generated using the MAP HOD. The left panel assumes $H_0 = 70$~km/s/Mpc and $\Omega_m = 0.274$, the fiducial cosmology assumed in \citet{2016Saito}, whereas the right panel assumes \citet{2016Planck} cosmology. The error bars are taken from the diagonal of the covariance matrices of the observables. We see that both the emulated $w_p$ and $\Delta \Sigma$ deviate significantly from observations. }
    \label{fig:emulated_genhod}
\end{figure*}

We first present the joint fit using the generalized HOD with the 6 parameters listed in the first column of Table~\ref{tab:prior_genhod}. We impose broad Gaussian priors centered around their baseline values, as listed in the second and third column of  Table~\ref{tab:prior_genhod}. The baseline values of all generalized parameters are set to 0. The width of the Gaussian priors are chosen to be broad and non-informative, but constraining enough to mostly fit inside the emulators' training range and reject unphysical values. 

\begin{table}
\centering
\begin{tabular}{ c | c c}
\hhline {===}
Parameter name & $\mu_{\textrm{prior}}$ & $\sigma_{\textrm{prior}}$ \\ 
\hline
$\log_{10}(M_{\textrm{cut}}/h^{-1}M_\odot)$ & 13.2 & 0.3  \\ 
$\log_{10}(M_1/h^{-1}M_\odot)$ & 14.2 & 0.3  \\
$\sigma$ & 0.897 & 0.1 \\
$\alpha$ & 1.151 & 0.1 \\
$s$ & 0 & 0.1 \\
$A$ & 0 & 0.1 \\
\hline 
\end{tabular} 

\caption{The prior information for the generalized HOD model. We choose the priors to be Gaussians centered on the baseline values with broad non-informative width. The baseline values are taken from \citet{2017bAlam}.}
\label{tab:prior_genhod}
\end{table}

Figure~\ref{fig:corner_genhod} shows the 1D and 2D marginalized posterior constraints on the generalized HOD parameters. The black lines show the baseline values as reference. The blue contours show the 0.5, 1, 1.5, and 2 $\sigma$ uncertainties. The values displayed above the 1D marginals are posterior medians with the upper/lower bounds associated with the 0.025 and 0.975 quantiles. Compared to the baseline values in Table~\ref{tab:baseline}, the posterior shows a preference for a lower $M_{\textrm{cut}}$, $\sigma$, and $A$, while the other parameters remain consistent with the prior mean.

We point out that, even though the posterior marginals look Gaussian, the posterior medians shown on the 1D marginals are different from the maximum a posteriori (MAP) point due to the complex shape of the posterior surface. We illustrate this point by showing an example trace plot for parameter $s$ in the top panel of Figure~\ref{fig:trace}, where we plot the posterior samples as a function of log prior volume, colored by their importance weight. The nested sampler marches from the left to the right and goes from being prior dominated to being likelihood dominated. The shape of the posterior marginal is determined by the most heavily weighted samples, colored in yellow, whereas the samples diverge into multiple local maxima towards the right of the plot. The MAP point corresponds to the long spike at $s \approx 0.15$, whereas the posterior median corresponds to the center of the yellow region, which is close to 0.

For completeness, the MAP point is given by $\log_{10}(M_{\textrm{cut}}/h^{-1}M_\odot) = 13.27$, $\log_{10}(M_1/h^{-1}M_\odot) = 14.34$, $\sigma = 0.634$, $\alpha = 1.144$, $s = -0.065$, and $A = -0.484$. Compared to the posterior median point shown on the 1D marginals, we see that the data actually prefer an even more extreme $A$. An assembly bias parameter of -0.5 is beyond our training range and also physically unlikely considering it would mean the shuffling of satellites across 1.0 dex in mass range, which is comparable to the entire halo mass range in the simulations. This fit suggests that the generalized HOD model failed to find a consensus between the 2PCF and g-g lensing signal within a reasonable parameter range.
%These values are probably unphysical considering that the best-fit $M_1$ (typical halo mass to host a satellite galaxy) is lower than the baseline value by an order of a magnitude. A 0.7 deviation in the value of $\alpha$ is also unlikely given that most previous results find $\alpha\sim 1$. As we will see shortly, we believe that this poor fitting heavior is caused by the clustering and lensing data sets being incompatible within this model space. 

The Bayesian evidence for this generalized HOD model is $\log \mathcal{Z} = -65.09 \pm 0.08$. The Bayesian evidence for the standard HOD model without any generalizations is $\log \mathcal{Z} = -69.46 \pm 0.07$. Thus, the generalized model is preferred over the standard model, with a Bayesian evidence roughly 4.4 e-fold higher than that of the standard model. Thus, the observed $w_p$ and $\Delta \Sigma$ somewhat favor the generalized HOD model despite the introduction of 2 new parameters, which increases the prior volume by order of a few. We summarize the evidence values for all the different models in Table~\ref{tab:logzs} for comparison. 

We show the maximum a posteriori (MAP) $w_p$ and $\Delta \Sigma$ in Figure~\ref{fig:emulated_genhod}. With nested sampling, the MAP is simply the last point in the chain. Again note that the MAP point is not the posterior median shown on the 1D marginal posterior plots. We then pass the MAP values back to the emulators to generate the emulated $w_p$ and $\Delta \Sigma$, which we plot in green. The red curves show the actual $w_p$ and $\Delta\Sigma$ computed from the mocks generated from the MAP HOD. We show the observed signals plotted in blue for comparison. It is clear that this generalized HOD model does not provide a good fit for either $w_p$ or $\Delta \Sigma$. Compare with \citet{2017Leauthaud}, which shows that the HOD that fits $w_p$ predicts a g-g lensing signal $20-40\%$ higher than the observed g-g lensing signal. This fit shows that we do not find a point in the generalized HOD space that fits both observables, but instead the model is trying to find a comprise between the observed $w_p$ and g-g lensing, by bringing down the predicted g-g lensing while losing consistency with the observed $w_p$. 

Comparing the emulated and mock observables in Figure~\ref{fig:emulated_genhod}, we see that our $\Delta\Sigma$ emulator successfully predicts the mock $\Delta\Sigma$ with remarkable accuracy. The $w_p$ emulator performs fairly well at $r_\perp > 1 h^{-1}$Mpc, but less so at smaller scales. This is due to the fact that the MAP point is somewhat outside of our training range due to an extreme best-fit assembly bias value. 

We also repeat this joint fit with prior widths on all the non-mass parameters extended from 0.1 to 0.2, allowing us to explore regions of the HOD parameter space beyond our training. We recover the same qualitative result. We do not find a consensus fit between the two observables even though the best fit HOD now gives more extreme values, with $\sigma = 0.369$ and $A = -0.848$. The best fit still significantly under-predicts $w_p$ while significantly over-predicting g-g lensing, similar to Figure~\ref{fig:emulated_genhod}.

\begin{figure*}
    \centering
    \hspace*{-0.7cm}
    \includegraphics[width=7.6in]{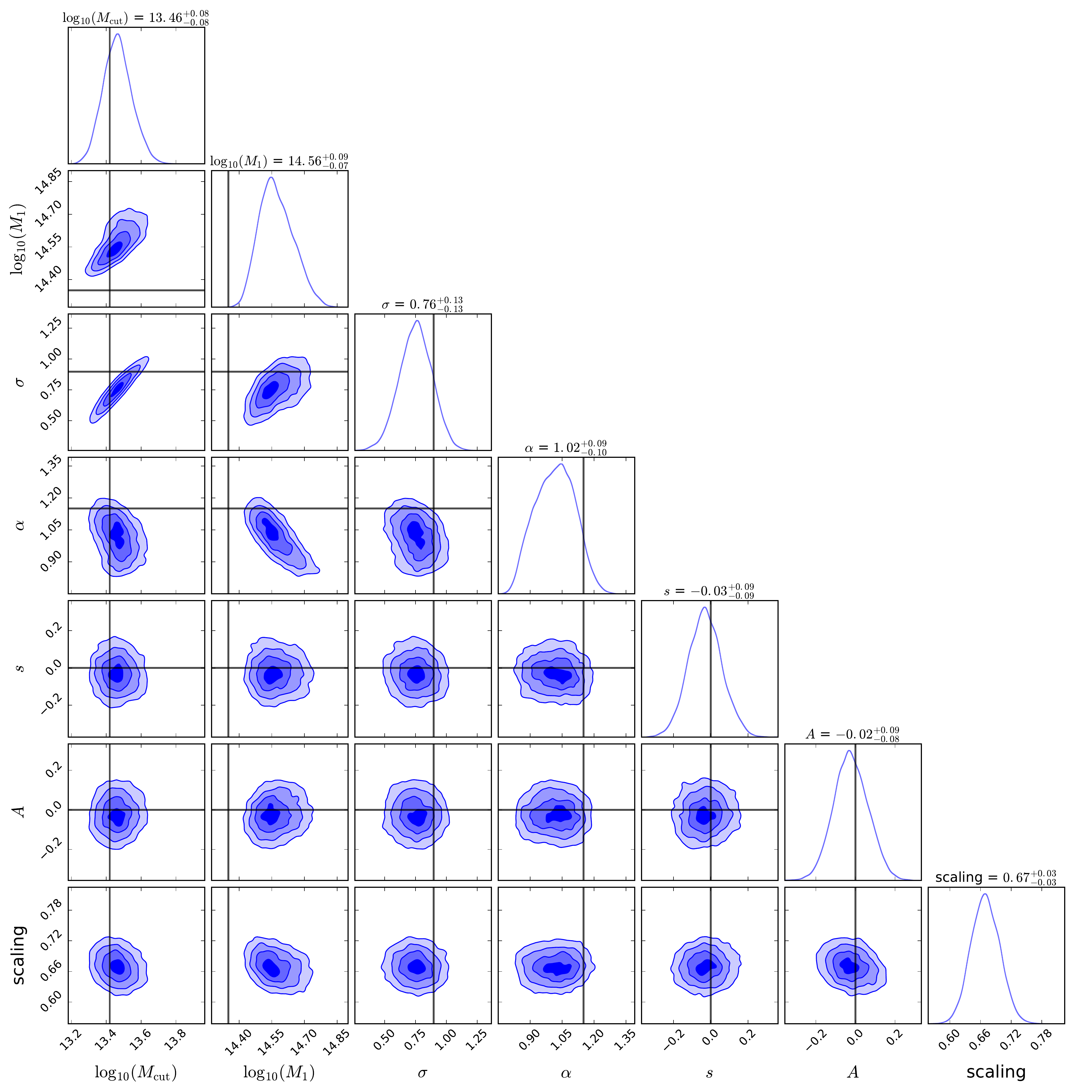}
    \vspace{-0.7cm}
    \caption{The 1D and 2D marginalized posterior constraints on the generalized HOD parameters, including a floating amplitude parameter ``scaling". The contours shown correspond to 0.5, 1, 1.5, and 2 $\sigma$ uncertainties. The vertical and horizontal lines show the centers of the Gaussian priors for reference. The values displayed above the 1D marginals are posterior medians with the upper/lower bounds associated with the 0.025 and 0.975 quantiles.}
    \label{fig:corner_genhod_scaling}
\end{figure*}

\begin{figure*}
    \centering
    \hspace*{-0.7cm}
    \includegraphics[width=7in]{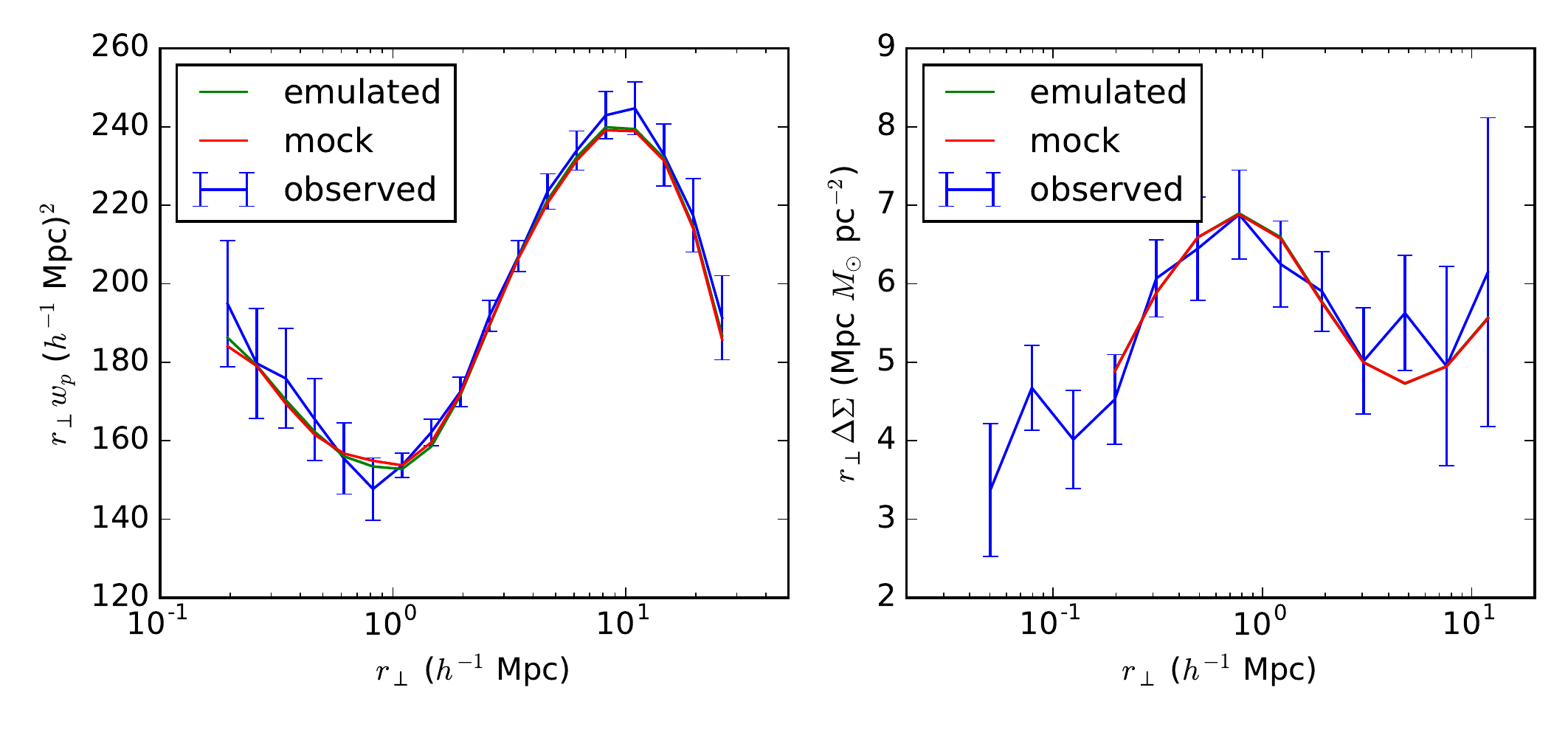}
    \vspace{-0.7cm}
    \caption{The maximum a posteriori (MAP) emulated $w_p$ and $\Delta \Sigma$ (in green) compared to the observed $w_p$ and $\Delta \Sigma$ (in blue) and the mock $w_p$ and $\Delta \Sigma$ (in red), using the 6-parameter generalized HOD model plus the ``scaling" parameter. The mock observables are computed from mock catalogs generated using the MAP HOD. The left panel assumes $H_0 = 70$~km/s/Mpc and $\Omega_m = 0.274$, the fiducial cosmology assumed in \citet{2016Saito}, whereas the right panel assumes \citet{2016Planck} cosmology. The error bars are taken from the diagonal of the covariance matrices of the observables. We see that both the emulated $w_p$ and $\Delta \Sigma$ are consistent with observations. }
    \label{fig:emulated_genhod_scaling}
\end{figure*}

We then test the 7-parameter generalized HOD model with the addition of satellite velocity bias $s_v$. The 7 parameters are $\log_{\textrm{10}} M_{\textrm{cut}}, \log_{\textrm{10}} M_1, \sigma, \alpha, s, s_v$, and $A$. We choose a broad Gaussian prior for $s_v$, with $\mu_{\textrm{prior}} = 0$ and $\sigma_{\textrm{prior}} = 0.1$.
It is possible that the addition of $s_v$ can improve the model, considering that the generalized HOD implementation of satellite velocity bias is based on particle selection so it does affect the radial distribution of the satellites in the halo. Figure~4 of \citet{2018Yuan} shows that $s_v = \pm 0.2$ changes $w_p$ by approximately $1\%$. 

We fit the 7-parameter generalized HOD and sample the posterior with \texttt{dynesty}. We find the same modes in the posterior space as the 6-parameter generalized HOD without $s_v$. The Bayesian evidence of this model is $\log \mathcal{Z} = -65.18 \pm  0.08$, which is statistically the same as that of the 6-parameter model despite the addition of a new parameter. Thus, the 7-parameter model with $s_v$ is not preferred over the 6-parameter model without $s_v$. We omit the $s_v$ parameter in the following models we consider. 
\subsection{Generalized HOD with an amplitude scaling parameter}

In this section, we introduce a floating amplitude parameter ``scaling" to the lensing signal. This is a free parameter multiplied onto the emulated $\Delta \Sigma$ to allow its overall amplitude to shift up and down. We set a log-normal prior on ``scaling'' with mode equal to 1 and a scale of 0.2. 

Figure~\ref{fig:corner_genhod_scaling} shows the 1D and 2D marginalized posterior constraints on the 6-parameter generalized HOD, but now with the ``scaling" parameter. The posterior median is less than 1$\sigma$ deviation in most HOD parameters from the prior mean, except for $\log_{10}M_1$, which gives a $\sim2\sigma$ deviation from prior mean. The MAP HOD is $\log_{10}(M_{\textrm{cut}}/h^{-1}M_\odot) = 13.36$, $\log_{10}(M_1/h^{-1}M_\odot) = 14.48$, $\sigma = 0.590$, $\alpha = 1.102$, $s = -0.221$, and $A = -0.098$, again somewhat different from the posterior medians shown on the 1D marginals. However, unlike the ``unscaled" fit, all the best-fit parameters of this fit fall inside the training range of our emulator, with moderate assembly bias and satellite distribution parameter values. The only parameter whose best-fit value is significantly different from the prior mean is $M_1$, approximately 0.28 dec higher than the baseline value, which calls for fewer satellites given a halo mass but this value is not beyond reason. The ``scaling" factor strongly favors a value of $\sim 0.67$ instead of 1. The Bayesian evidence of this model is $\log\mathcal{Z} = 16.00 \pm  0.08$, a 49.1 e-fold increase over the 6-parameter model with no floating amplitude. Thus, the inclusion of a floating amplitude parameter for the g-g lensing signal is strongly favored and we now find a reasonable consensus HOD between the two observables. 

\begin{figure}
    \centering
    \hspace*{-0.6cm}
    \includegraphics[width=3.7in]{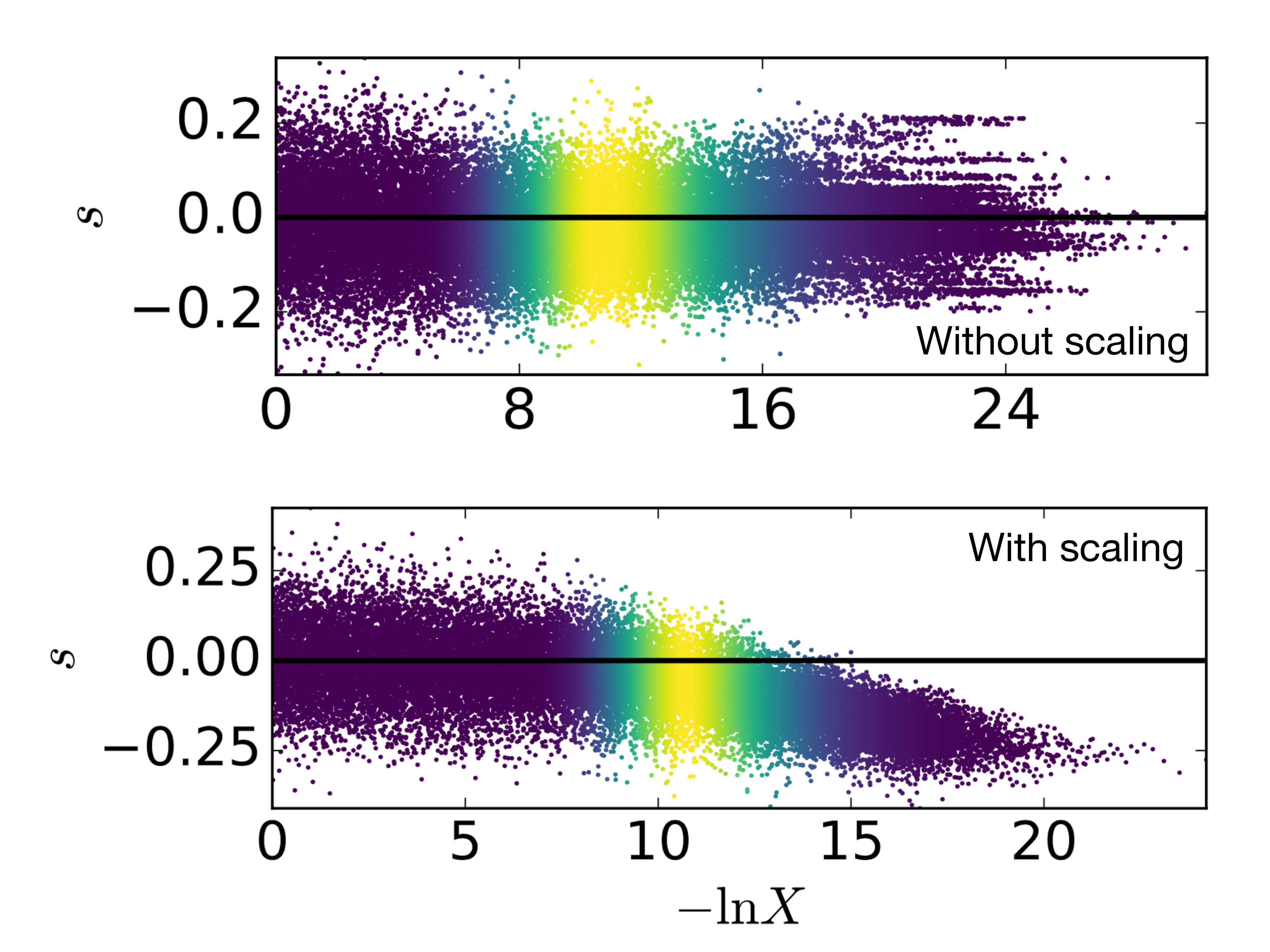}
    \vspace{-0.3cm}
    \caption{The trace plot of the $s$ parameter, showing 1-D positions of samples (dead points) over the course of the run, colored by their estimated importance
weight PDF $p(X)$. The top and bottom panels correspond to the fits without and with the scaling parameter, respectively. The $x$ axis denotes log prior volume. }
    \label{fig:trace}
\end{figure}

We show the trace plot for parameter $s$ in Figure~\ref{fig:trace}, where the top and bottom panels show the trace plots for the fit without and with the scaling parameter, respectively. The nested sampler marches from the left to the right through the prior volume. The points denote the dead points over the course of the run, colored by their importance weights. For technical details, refer to the \texttt{dynesty} code release paper \citep{2019Speagle}. We see that for both runs, the sampling is initially prior dominated but becomes more and more likelihood dominated as we move to the right of the panel. For the ``unscaled" case, the sample diverges and several local modes develop, suggesting disagreement in the dataset. However, for the ``scaling" case, the sample remains compact throughout the prior volume, giving rise to one maximum likelihood mode, suggesting the discovery of a consistent fit.

Figure~\ref{fig:emulated_genhod_scaling} shows the emulated and mock $w_p$ and $\Delta\Sigma$ of the MAP estimate of the 6-parameter + scaling model posterior. The emulated $\Delta\Sigma$ shown is after applying the best-fit scaling parameter, thus appearing in excellent agreement with the observed $\Delta\Sigma$. We see that both the $w_p$ and $\Delta\Sigma$ predictions are consistent with the observations, deviating by $<1\sigma$ in almost all bins. Unlike for the unscaled fit, the emulated observables and the mock observables are also in excellent agreement on all scales, showcasing the high accuracy of our emulators within its training range. The fact that the introduction of a flexible amplitude on $\Delta\Sigma$ produces a remarkably good fit with reasonable best-fit HOD parameters suggests that the inconsistency between the observed $w_p$ and $\Delta\Sigma$ measurements is well described by a scale independent amplitude shift. The best-fit value of 0.67 for the scaling parameter corresponds to a $33\%$ deficiency in bias, which is consistent with the $20-40\%$ disagreement between the observed and predicted $\Delta\Sigma$ found in \citet{2017Leauthaud}. This suggests that assembly bias alone cannot reconcile the observed $w_p$ and $\Delta\Sigma$ measurements. 

\subsection{Joint fits with decorated HOD}
\label{subsec:fits_decHOD}

The decorated HOD implementation provided in \texttt{Halotools} provides another framework to extend the standard 5-parameter HOD model. The decorated HOD presents several key differences compared to the generalized HOD. The decorated parameters include two separate assembly bias parameters for the central and satellite galaxies. A key difference in implementation is that it distributes satellite galaxies on NFW profiles given halo concentration whereas the generalized HOD distributes the satellites on halo particles, leading to differing observable predictions on the small scale. While we showed that the generalized HOD in its current form does not reconcile the observed projected clustering and g-g lensing signal, it is possible that the decorated HOD model provides a better joint fit of the two. 

\begin{table}
\centering
\begin{tabular}{ c | c c}
\hhline {===}
Parameter name & $\mu_{\textrm{prior}}$ & $\sigma_{\textrm{prior}}$ \\ 
\hline
$\log_{10}(M_{\textrm{cut}}[h^{-1}M_\odot])$ & 13.2 & 0.3 (truncated at $\pm1$) \\ 
$\log_{10}(M_1[h^{-1}M_\odot])$ & 14.2 & 0.3 (truncated at $\pm1$) \\
$\sigma$ & 0.897 & 0.1 \\
$\alpha$ & 1.151 & 0.1 \\
$A_{\textrm{cent}}$ & 0 & 0.1 \\
$A_{\textrm{sat}}$ & 0 & 0.1 \\
\hline 
\end{tabular} 

\caption{The prior specification for the decorated HOD model. We choose the priors to be Gaussians centered on the baseline values with broad non-informative width.}
\label{tab:prior_dechod}
\end{table}

% \begin{figure*}
%     \centering
%     \hspace*{-0.7cm}
%     \includegraphics[width=7.6in]{./cornerplot_mapstart_halotools_selected_all2nds_dynesty_standard_large_20_1500.pdf}
%     \vspace{-0.7cm}
%     \caption{The 1D and 2D marginalized posterior constraints on the decorated HOD parameters, which include the 4 standard HOD parameters and assembly bias parameters $A_{\textrm{cent}}$ and $A_{\textrm{sat}}$. The contours shown correspond to 0.5, 1, 1.5, and 2 $\sigma$ uncertainties. The vertical and horizontal lines show the centers of the Gaussian priors for reference. The values displayed above the 1D marginals are posterior medians with the upper/lower bounds associated with the 0.025 and 0.975 quantiles.}
%     \label{fig:corner_decHOD}
% \end{figure*}
\begin{figure*}
\centering
    \subfigure[]{    
    \hspace*{-0.7cm}
    \includegraphics[width=7in]{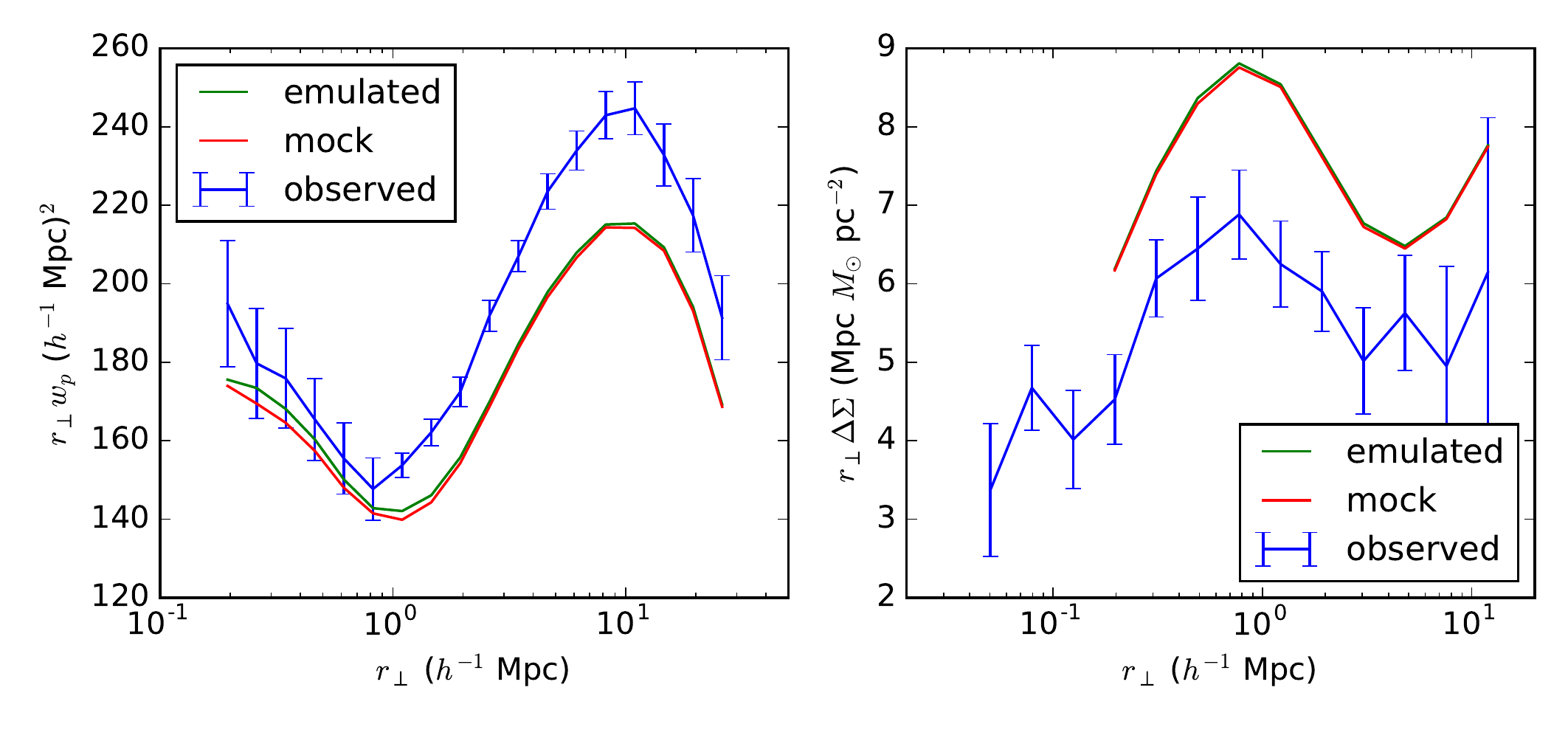}
    \label{fig:emulated_dechod}
    \vspace{-0.7cm}
    }
    \subfigure[]{    
    \hspace*{-0.7cm}
    \includegraphics[width=7in]{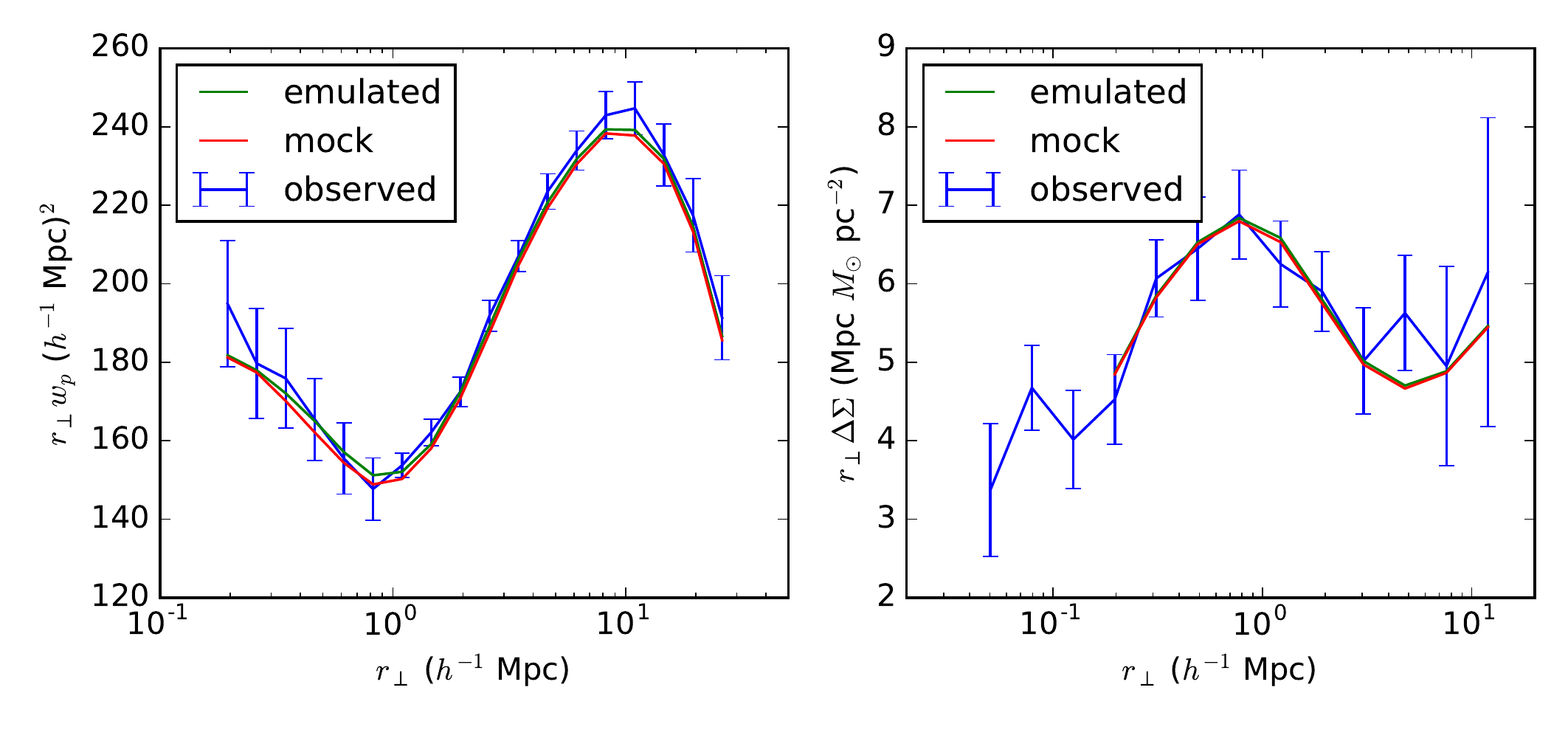}
    \label{fig:emulated_dechod_scaling}
    \vspace{-0.7cm}
    }
    \caption{(a) The maximum a posteriori (MAP) emulated $w_p$ and $\Delta \Sigma$ (in green) compared to the observed $w_p$ and $\Delta \Sigma$ (in blue) and the mock $w_p$ and $\Delta \Sigma$ (in red), using the decorated HOD model with $A_\textrm{cent}$ and $A_\textrm{sat}$, without the scaling parameter. (b) The same as (a), but with the scaling parameter. We see that without the scaling parameter, the best-fit emulated $w_p$ and g-g lensing are inconsistent with observations. With the scaling parameter, both the emulated $w_p$ and $\Delta \Sigma$ are in excellent agreement with the observations.
    The left panels assume $H_0 = 70$~km/s/Mpc and $\Omega_m = 0.274$, the fiducial cosmology assumed in \citet{2016Saito}, whereas the right panels assume \citet{2016Planck} cosmology. The error bars are taken from the diagonal of the covariance matrices of the observables.}
    \label{fig:emulated_dechod_both}
\end{figure*}

% Figure~\ref{fig:corner_decHOD} shows the 1D and 2D marginalized posterior constraints on the decorated HOD parameters. The black lines show the baseline values as reference. The blue contours show the 0.5, 1, 1.5, and 2 $\sigma$ uncertainties. The parameter values quoted above the 1D marginals are the medians of the posterior with the upper/lower bounds corresponding to the 0.025 and 0.975 quantiles.
The construction of the decorated HOD emulator was discussed in Section~\ref{subsec:model}. Then we use \texttt{dynesty} to perform a joint fit of the observed $w_p$ and g-g lensing following the same procedure as for the generalized HOD framework. The 6 HOD parameters in this joint fit are $\log_{\textrm{10}} M_{\textrm{cut}}, \log_{\textrm{10}} M_1, \sigma, \alpha, A_{\textrm{cent}}$, and $A_{\textrm{sat}}$. We again choose broad non-informative Gaussian priors for these parameters centered around their baseline values. However, we truncate the Gaussians at -1 and 1 for the two assembly bias parameters, as they are only defined between -1 and 1. We summarize these priors in Table~\ref{tab:prior_dechod}.

The MAP HOD values are $\log_{10}(M_{\textrm{cut}}/h^{-1}M_\odot) = 13.01$, $\log_{10}(M_1/h^{-1}M_\odot) = 14.10$, $\sigma = 0.562$, $\alpha = 1.299$, $A_{\textrm{cent}} = -0.168$, and $A_{\textrm{sat}} = -0.387$. We see that the data prefer a very high satellite assembly bias value of almost $-0.4$, which is likely unphysical as it suggests the less concentrated half of halos have more than twice as many galaxies as the more concentrated half. Figure~\ref{fig:emulated_dechod} shows the emulated $w_p$ and g-g lensing of the MAP HOD using the 6 parameter decorated HOD framework. We do not find a good fit, and we see behavior very similar to Figure~\ref{fig:emulated_genhod} in that the joint fit attempts to bring the emulated lensing signal in better agreement with the observations by shifting the emulated $w_p$ out of agreement with the observations. The trace plot for this fit shows strong multi-modality, again suggesting that the model failed to find a consensus between the two observables within a reasonable parameter range. We do not show it for brevity.
% We see a strong multi-modality that shows the complex structure of the posterior space. We see strong correlation between $M_{\textrm{cut}}$, $M_1$, and $\sigma$.
% The MAP mode favors a higher $M_\textrm{cut}$, $M_1$, $\sigma$, and $\alpha$, but a strong negative central assembly bias and a negative satellite assembly bias. 
% Compared to Figure~\ref{fig:corner_genhod}, the decorated HOD fit recovers rather different best-fit HOD parameters, notably favoring values on the opposite side of the baseline values shown in black lines. The 4 standard HOD parameters' best-fit values are all within $3\sigma$ of the baseline values shown in black lines. However, the central assembly bias $A_{\textrm{cent}}$ does deviate significantly from 0, and a value of $\sim 0.8$ suggests that it is possibly limited by the prior, which is a Gaussian centered on 0 with a width of 0.3. It is possible that an even more extreme $A_{\textrm{cent}}$ is favored with a broader prior, but keep in mind that $-1 < A_{\textrm{cent}} < 1$ by construction. 

The evidence of this model is given by $\log \mathcal{Z} = -72.89 \pm 0.08$, compared to the evidence of the standard 5-parameter model without the 2 assembly bias parameters $\log\mathcal{Z} = -74.91 \pm  0.07$. This suggests that the decorated HOD model with assembly bias is moderately favored over the standard 5-parameter HOD. The evidence is also 7.8 e-fold lower than that of the generalized HOD model, possibly due to the inclusion of satellite distribution parameter $s$ and differences in the implementation. 
Again we have compiled all the evidence values in Table~\ref{tab:logzs} for comparison.

% Figure~\ref{fig:emulated_dechod} shows the emulated $w_p$ and $\Delta\Sigma$ of the MAP estimate of the decorated HOD parameter posterior. Similar to what we see for the best-fit of the generalized HOD model in Figure~\ref{fig:emulated_genhod}, the emulated signals are inconsistent with the observed signals by as much as $4\sigma$. Together with the low integrated evidence, this suggests the inclusion of a more flexible assembly bias model with separate assembly bias dependencies for the central and satellite galaxies does not remedy the inconsistencies between the observed projected clustering and g-g lensing measurements. 

%\al{Move this new section 5.3. It will make the paper flow better.}

%\al{Also, At the top of 5.3, can we say add a little more motivation for this scaling parameter. Why are we introducing this? What are we testing? It doesnt come across clearly in the current text} 

Following the same procedure as for the generalized HOD, we then add the scaling parameter to the decorated HOD model to allow the overall amplitude of the lensing signal to shift. We re-fit the observed $w_p$ and $\Delta\Sigma$ and find a set of best-fit values that seem more reasonable and give emulated signals that closely match the observables. The best-fit HOD parameters are $\log_{10}(M_{\textrm{cut}}/h^{-1}M_\odot) = 13.19$, $\log_{10}(M_1/h^{-1}M_\odot) = 14.28$, $\sigma = 0.601$, $\alpha = 1.370$, $A_{\textrm{cent}} = 0.136$, and $A_{\textrm{sat}} = -0.079$. The best-fit value for the scaling parameter is $0.66^{+0.03}_{-0.03}$, which represents a $34\%$ inconsistency between the observed $w_p$ signal and the observed $\Delta\Sigma$ signal, consistent with the discrepancy reported in \citet{2017Leauthaud}. We show the best-fit emulated $w_p$ and g-g lensing signal in Figure~\ref{fig:emulated_dechod_scaling}, where we see excellent agreement between the emulated signals and the observed signals. The integrated Bayesian bias for this model is given by $\log\mathcal{Z} = -17.92\pm0.08$, which is a 55 e-fold increase compared to the decorated model without scaling. This again shows that the inclusion of a flexible lensing amplitude is strongly favored, and that the decorated HOD model by itself does not provide a good fit of the projected clustering and g-g lensing observables. We recover the same qualitative results when we increase the priors on the non-mass parameters ($\sigma$, $\alpha$, $A_\textrm{cent}$, and $A_\textrm{sat}$) from 0.1 to 0.2, though the joint fit in the unscaled case returns more extreme best-fit values in $\sigma$ and the assembly biases.

It is worth noting that Figure~\ref{fig:emulated_dechod_both} shows great consistency between the emulated $w_p$ in green and the mock $w_p$ in red, except at smaller scales for the unscaled case, where we get approximately $2\%$ error. We achieve such good accuracy due to our large training range and restricting the joint fit to the training range. When we increase the HOD parameter priors and let the joint fit explore regions that the emulators are not trained in, the accuracy quickly collapses, especially in $w_p$ due to its highly non-linear dependence on HOD parameters. We see a minor manifestation of this in Figure~\ref{fig:emulated_genhod}, where the best-fit assembly bias value is moderately beyond the training range.

It is also interesting that while the two HOD frameworks produce the same qualitative behavior in both the scaled and unscaled case and noticeably yielding the same best-fit scaling parameter, 
the best-fit decorated HOD and the best-fit generalized HOD give rather different HOD parameter values. This is possibly due to differences in the base HOD implementation, the assembly bias implementation, and the inclusion of the satellite distribution parameter $s$ in the generalized HOD. This also showcases the effect of differences in HOD implementations and highlights the importance of comparing different HOD models in future cosmological inference work.

\begin{table}
\centering

\begin{tabular}{ c | c c}
\hhline {===}
 & No scaling & scaling \\ 
\hline
generalized HOD (standard) & -69.46$\pm$0.07 &  \\ 
generalized HOD ($s$, $A$) & -65.09$\pm$0.08 & -16.00$\pm$0.08  \\ 
generalized HOD ($s$, $A$, $s_v$) & -65.18$\pm$0.08 & -16.20$\pm$0.08  \\ 
\hline 
decorated HOD (standard) & -74.91$\pm$0.07 &  \\ 
decorated HOD ($A_\textrm{cent}$, $A_{\textrm{sat}}$) & -72.89$\pm$0.08 & -17.92$\pm$0.08  \\ 
\hline 
\end{tabular} 

\caption{Summary of the evidence $\log\mathcal{Z}$ and their uncertainties of the different models we tested. The two columns correspond to the model without a scaling parameter and with a scaling parameter. The first three rows show the generalized HOD implementation in Section~\ref{subsec:fits_genHOD}, with the first row showing the standard 5-parameter model without any generalized parameter, and the next two showing the standard model plus the parameters in parentheses. The next two rows show the evidence of the decorated HOD implementation provided in \texttt{Halotools}.}
\label{tab:logzs}
\end{table}

\section{Discussion}
\label{sec:discussions}

In this paper, we have tested whether two different extended HOD models can fit the observed projected clustering signal $w_p$ and the g-g lensing signal $\Delta\Sigma$. We find that neither model provides a good joint fit, with Bayesian evidence more than 49 e-fold lower than that of the models with an extra scaling parameter. The best-fit scaling parameter is consistently showing a roughly $34\%$ discrepancy between the observed $w_p$ and $\Delta \Sigma$, in agreement with the conclusions of \citet{2017Leauthaud}.

Table~\ref{tab:logzs} summarizes the integrated evidence of the models we have tested in this paper. We see that in both the generalized HOD framework and the decorated HOD framework, the addition of generalized parameters --- $s, A$ for the generalized HOD and $A_{\textrm{cent}}, A_{\textrm{sat}}$ for the decorated HOD --- is favored, with a 4.3 and 2.0 e-fold increase to the model evidence, respectively. The addition of satellite velocity bias parameter is dis-favored, however, with no significant change to the model evidence. 
The decorated HOD models also show significantly lower evidence than the generalized HOD models. This is due to different implementations of the standard HOD, different implementations of assembly bias, and the inclusion of a satellite distribution parameter $s$ in the generalized HOD that introduces additional flexibility. 

The fact that the joint fit with generalized/decorated parameters give complex and multi-modal posteriors (see Figure~\ref{fig:corner_genhod}) are most likely a result of the clustering and lensing data sets being incompatible, but it may also suggest that our HOD parameter space has moderate degeneracies with respect to the projected clustering and g-g lensing data sets. 
It is possible that if we fit to the anisotropic correlation function $\xi(r_\perp, \pi)$ instead of the $w_p$, we would recover more constraining power on the parameter posteriors. 

Our conclusions are limited in several ways. First of all, we only considered galaxy assembly bias implementation that uses halo concentration as the secondary dependency. In the more general sense of the term, galaxy assembly bias is not just limited to  halo concentration as the only secondary dependence. It is possible that a galaxy assembly bias implementation whose secondary dependence is linked to the merger history, or local environment, or other halo properties, may reconcile the observed discrepancy. 
\citet{2019Lange} additionally tested the decorated HOD model with halo spin as the secondary dependence and found a decrease of at most $10\%$ to the predicted lensing signal on the small scale and almost no impact on the large scale, insufficient to reconcile the $20-40\%$ discrepancy.

We have encountered significant challenges when constructing our $w_p$ emulator over a wide range of HOD parameters. We find that $w_p$ has a highly non-linear dependence on the HOD parameters, making the second-order Taylor expansion model insufficient in modeling $w_p$ over the full range of HOD parameters. Thus, we implement the rejection sampling step to only emulate in regions of HOD space that are well-behaved. Even then, we still have difficulties in the accuracy of our decorated HOD emulator in certain regions of parameter space that are of potential interest. In this work, we have limited ourselves to exploring a large but not exhaustive range of HOD parameters. In the future, we intend to further expand our training set and adopt a neural network style emulator that can more easily incorporate these non-linearities. (Refer to \citet{2019Wibking} for a non-linear emulator implementation using Gaussian processes.) We also plan on changing to a different HOD parameter basis, possibly using parameters such as total galaxy number density and satellite fraction, to reduce degeneracies in the parameters.

Another limitation of our analysis is that we do not marginalize over Planck cosmological posteriors. We used a fixed cosmology in building our emulators and our fits. We believe the cosmological dependence of this problem is an interesting one, and in future work we will use the \textsc{AbacusCosmos} simulation boxes that sample the Planck posterior to extend our emulator to also emulating the cosmological parameters. We defer that discussion to a future paper. 

Baryonic effects provide another possible explanation of the discrepancy between the projected clustering measurements and the g-g lensing measurements. We have used dark matter only N-body simulations for our model predictions. While the introduction of the satellite distribution parameter may have made our generalized HOD models flexible enough to marginalize over baryonic effects in the projected clustering predictions, it does not change the distribution of dark matter particles themselves and thus fails to account for any baryonic effects on the g-g lensing signal. Several studies have suggested that baryonic effect can impact the halo profile and influence subhalo properties \citep[e.g.][]{2014vanDaalen, 2014Velliscig, 2016Chaves}. \citet{2017Leauthaud} compared the g-g lensing signals from the full-physics Illustris simulations \citep{2014aVogelsberger, 2014bVogelsberger, 2014Genel, 2015Sijacki, 2015Nelson} and from the corresponding gravity-only simulations to estimate the effect of baryonic physics on CMASS-like samples, and found that baryonic effects can induce a noticeable increase in $\Delta\Sigma$, in the direction of reconciling the $20-40\%$ discrepancy. Refer to Figure~12 of \citet{2017Leauthaud}, we see that baryonic physics can induce an increase of $\sim 20\%$ in $\Delta\Sigma$ at a radius of 0.2$h^{-1}$Mpc, and an essentially zero increase at larger radius of approximately $10 h^{-1}$Mpc. We find in our analyses that the $33-34\%$ discrepancy in the lensing signal exists on all scales, so baryonic effects, at least as detected using the Illustris simulations, might not fully explain the lensing discrepancy. Moreover, \citet{2017Weinberger} found a smaller impact of the baryonic effects using an improved AGN feedback model with the Illustris simulations. 
\citet{2019Lange} used the newly-released and improved IllustrisTNG simulation \citep{2018Pillepich, 2018Springel, 2019Nelson, 2018Naiman, 2018Marinacci, 2018Nelson} and found an at most $10\%$ decrease to the lensing signal on the small scale due to baryonic effects, even in the $95\%$ posterior range.
However, beyond just affecting the halo density profile, baryonic effects can also produce more complicated assembly bias beyond that using the halo concentration proxy.

\section{Summary and Conclusions}
\label{sec:conclusions}

In this paper, we address the question of whether galaxy assembly bias, defined as the secondary dependence of galaxy occupation on halo concentration, can explain the $20-40\%$ discrepancy between the galaxy projected clustering measurement $w_p$ and the g-g lensing measurement $\Delta \Sigma$ in a Planck cosmology. We apply two different extensions to the standard 5-parameter HOD of the \citet{2007Zheng} form to fit the clustering and lensing observables. The first is a generalized HOD which adds the satellite distribution parameter and the assembly bias parameter. The second is the decorated HOD, which adds the central assembly bias parameter and the satellite assembly bias parameter. We find that neither model can reconcile the discrepancy between the two observables. However, allowing a $\sim34\%$ increase to the measured lensing amplitude would yield a very good fit using either models. This result suggests that galaxy assembly bias, in the ways we have implemented it, does not reconcile the $20-40\%$ discrepancy between the galaxy projected clustering measurement and the g-g lensing measurement in Planck cosmology. While other implementations of galaxy assembly bias might explain the discrepancy, our findings suggest that galaxy assembly bias is in-plausible as the main explanation of the discrepancy. It is also possible that a slightly different cosmology or baryonic effects can explain the discrepancy, but these discussions are beyond the scope of this paper.

\section*{Acknowledgements}

We thank our referee Andrew Hearin for helpful feedback on this paper. We also thank him and Kuan Wang for help with \texttt{Halotools}. We would also like to thank Josh Speagle and Lehman Garrison for their support with \texttt{dynesty} and \textsc{AbacusCosmos} throughout this paper. We are grateful to Shun Saito for allowing us to use their projected clustering measurement data. 

DJE is supported by U.S. Department of Energy grant DE-SC0013718 and as a Simons Foundation Investigator. AL is supported by the U.D Department of Energy, Office of Science, Office of High Energy Physics under Award Number DE-SC0019301, by the Packard Foundation, and by the Sloan Foundation.

The \textsc{ABACUS} simulations used in this paper are available at \url{https://lgarrison.github.io/AbacusCosmos}.
\bibliographystyle{aasjournal}
\bibliography{biblio}
\end{document}